\RequirePackage{fix-cm}
\documentclass[smallextended]{svjour3}       % onecolumn (second format)
\smartqed  % flush right qed marks, e.g. at end of proof
\usepackage{graphicx}
\usepackage[]{fontenc}
\usepackage[latin1]{inputenc}
\usepackage{verbatim}
\usepackage{amsmath}
\usepackage{amssymb}
\usepackage{dsfont}
\usepackage{amsfonts}
\usepackage{mathrsfs}
\usepackage{graphicx}
\usepackage{color}
\usepackage{ulem}
\usepackage{xspace,bbm,epic,eepic}
\usepackage{pgfplots}
\usepackage{setspace}
\usepackage{xspace}

\usepackage{ stmaryrd }
\usepackage{graphicx}
\usepackage{lineno}
\usepackage{color}
\usepackage{url}
\usepackage{graphicx}
\usepackage{ stmaryrd }
\usepackage{soul}		         % utilisÃ© dans les macros de Charles

\usepackage{amsmath,amssymb,color} % define this before the line numbering.
\newcommand{\s}{\mathbf{s}}
\newcommand{\w}{w}

\newcommand{\Rr}{\mathds{R}}

\newcommand{\D}{\mathbf{u}}

\newcommand{\R}{\mathbf{p}}

\newcommand{\X}{\mathbf{X}}\newcommand{\x}{\mathbf{x}}
\newcommand{\Y}{\mathbf{Y}}\newcommand{\y}{\mathbf{y}}
\newcommand{\XI}{\boldsymbol{\xi}}

\newcommand{\eg}{\textit{e.g.}, }
\newcommand{\ie}{\textit{i.e.}, }

\DeclareMathOperator*{\argmin}{arg\,min}

\def\paramSet{\Theta}
\def\sparamSet{\tilde{\paramSet}}
\def\param{\theta}
\def\px{p_\X}
\def\py{p_\Y}
\def\pyx{p_{\Y\vert\X}}
\def\pxy{p_{\X\vert\Y}}

\def\spx{\tilde{p}_\X}
\def\sp{\tilde{p}}
\def\ppx{\hat{p}_\X}
\def\pp{\hat{p}}
\def\ppy{\hat{p}_\Y}
\def\ppxy{\hat{p}_{\X\vert\Y}}
\def\solman{\mathcal{M}}
\def\obs{{\y}}
\def\state{{\x}}
\def\statea{{\tilde{\x}}}

\begin{document}

\title{Reduced Modeling of Unknown Trajectories%\thanks{Grants or other notes
%about the article that should go on the front page should be
%placed here. General acknowledgments should be placed at the end of the article.}
}
%\subtitle{Do you have a subtitle?\\ If so, write it here}

%\titlerunning{Short form of title}        % if too long for running head

\author{ Patrick H\'eas, C\'edric Herzet
}

%\authorrunning{Short form of author list} % if too long for running head

\institute{P. H\'eas \at
             INRIA Centre de Rennes - Bretagne Atlantique, campus universitaire de Beaulieu, 35042 Rennes, France  \\
              Tel.: +33 2 99 84 75 50\\
              \email{patrick.heas@inria.fr}           %  \\
%             \emph{Present address:} of F. Author  %  if needed
           \and
           C. Herzet \at
              INRIA Centre de Rennes - Bretagne Atlantique, campus universitaire de Beaulieu, 35042 Rennes, France
}
%\institute{}
\date{
%Received: date / Accepted: date
}
% The correct dates will be entered by the editor

\maketitle

\begin{acknowledgements}
{This work was
supported by the  ``Agence Nationale de la Recherche" (ANR) through the GERONIMO project.
 }
\end{acknowledgements}

\begin{abstract}
{This paper deals with model order reduction of parametrical dynamical systems. We consider the specific setup where the distribution of the system's trajectories is unknown but the following two sources of information are available: \textit{(i)} some ``rough'' prior knowledge on the system's realisations; \textit{(ii)} a set of ``incomplete'' observations of the system's trajectories.~We propose a Bayesian methodological framework to build reduced-order models (ROMs) by exploiting these two sources of information. 
 We emphasise that complementing the prior knowledge with the collected data provably enhances the knowledge of the distribution of the system's trajectories. 
We then propose an implementation of the proposed methodology based on Monte-Carlo methods. In this context, we show that standard ROM learning techniques, such \eg Proper Orthogonal Decomposition or Dynamic Mode Decomposition, can be revisited and recast within the probabilistic framework 
considered in this paper.~We illustrate the performance of the proposed approach by numerical results obtained for a standard geophysical model.}% show the gain brought by exploiting 
%\keywords{First keyword \and Second keyword \and More}
% \PACS{PACS code1 \and PACS code2 \and more}
% \subclass{MSC code1 \and MSC code2 \and more}
\end{abstract}

\section{Introduction}
%We introduce in what follows essential elements for the understanding of the paper's problematic.
%\remCH{J'enleve toutes les sous-sections, ca fait beaucoup trop compartiment\'e (2 niveaux de sous sections dans l'intro...)}

In many fields of Sciences, one is interested in studying the {spatio-temporal} evolution of a state variable characterised by  a  differential equation.  Numerical discretisation in space and time leads to parameterised high-dimensional systems of equations {of the form:} \vspace{-0.cm}
\begin{align}\label{eq:model_init} 
 \left\{\begin{aligned}
& x_{t}= f_t(x_{t-1},\theta) , \\%\mbox{\quad$\forall\, s\in\Sc,t\in\Tc$},
&x_1=g({\theta}),
\end{aligned}\right. \vspace{-0.cm}%\\
\end{align} 
where $x_t\in \Rr^n$ is the state variable,  ${\theta} \in \paramSet$ denotes some  parameters and $f_t:\Rr^n \times \paramSet  \to  \Rr^n$, $g:\paramSet \to \Rr^n$.
%As a few examples of systems obeying this type of constraints, we can mention the wave equation characterising the propagation of sound  \cite{Godlewski96}, the Navier-Stokes equations describing fluid evolution \cite{quartapelle13} or the Maxwell's equations governing the realm of electromagnetism \cite{Godlewski96}.
 {Because \eqref{eq:model_init} may correspond to very high-dimensional systems, computing a trajectory $\{x_t\}_{t=1}^T$ 
  may  lead to unacceptable computational burdens in some applications.
 %As a response to this computational bottleneck,  reduced-order models (ROMs) aim at approximating  the trajectories of the system for a range of regimes determined by a set of parameters \cite{2015arXiv150206797C}.  

 As a response to this bottleneck,  reduced-order models (ROMs) aim at providing ``good'' approximations of the trajectories of \eqref{eq:model_init} (in some particular regimes of interest) via  strategies only requiring  significantly-reduced computational resources.
 %~\remCH{Je pense cette phrase et la suivante n'apporte pas grand chose pour la suite et peuvent etre enlevee.} A common approach consists of assuming that the trajectories of interest are well approximated in some subspace $\approxss$ of $\Rr^n$. In this spirit, many  tractable low-rank approximations \remCH{"low-rank approximation" ne veut pas dire grand chose ici} of  high-dimensional systems have been proposed in the literature. 
 Among the most familiar reduction techniques, let us mention Galerkin projection using proper orthogonal decomposition (POD)~\cite{9780511622700} or reduced basis~\cite{Quarteroni2011Certified}, low-rank dynamic mode decomposition (DMD)~\cite{Chen12,HeasHerzet16,Jovanovic12},  second-order nonlinear operator approximation~\cite{peherstorfer2016data},  balanced truncation~\cite{Antoulas2005Overview} or Taylor expansion~\cite{ZAMM:ZAMM19830630105}. 

All the techniques  mentioned above presuppose (explicitly or implicitly) the knowledge of the  trajectories that the ROM should accurately approximate.  In many contributions, such a knowledge is characterised by the (so-called) solution manifold defined as
\begin{align}\label{eq:solman}
\solman = \{ \x = (x_1 \cdots x_T) \in \Rr^{n \times T} : \mbox{$\x$ obeys \eqref{eq:model_init} for some $\theta\in\paramSet$} \},
\end{align}
%where $\paramSet$ is some set of parameters, 
see \eg {\cite{2015arXiv150206797C}}. In this paper, we consider a slightly more general setting by assuming that the set of trajectories to reduce are specified by a probability density on $\x$, say $\px$.\footnote{The latter density can for example be defined via \eqref{eq:model_init} by imposing a probability density on $\theta\in\paramSet$.} Unlike the standard formulation \eqref{eq:solman}, density $\px$ then provides information on both the set of trajectories of interest (which corresponds to $\px\neq 0$) and their probability of occurence. 

%In this paper, letting $\x = (x_1 \cdots x_T) \in \Rr^{n \times T}$, we will assume that such a knowledge takes the form of a probability density on $\x$, say $\px$.\footnote{}

%They are  generated by varying parameter $\theta$ over an admissible set $\paramSet$. 
%%\begin{align*}
%%\mathcal{M}= \{\x:  \x  \mbox{ obeys \eqref{eq:model_init} with $\theta\in \paramSet$ }  \},
%%\end{align*}
%%where $\x = (x_1 \cdots x_T) \in \Rr^{n \times T}$. 
%We  consider the general setting where $\theta\in \paramSet$ is the realisation of a random variable $\Theta$, and therefore, according to   \eqref{eq:model_init}, we  assume that   $\x = (x_1 \cdots x_T) \in \Rr^{n \times T}$ is the realisation of a random variable $\X$ of \textit{prior}  probability distribution  denoted by $p_\X$. %In consequence, to each subset $d\x$ of the manifold  $ \mathcal{M}$  will correspond a probabilistic  measure $p_\X(d\x)$.
% %\remCH{Mentionner ici la littÃrature sur la sÃlection de snapshots du manifold $\mathcal{\tilde M}$.} 
%%
%%This manifold  is generated by a set of  admissible parameters $\paramSet\subset \Rr^p$ of interest.
%% In words,  
%% the manifold of interest $\mathcal{M}$ is defined by the set $\paramSet$ of admissible parameters, which determines  state trajectories through recursion  \eqref{eq:model_init}.  

Unfortunately, in practice a precise knowledge of $\px$ is usually not available. In this paper, we thus address the following question: how to build a good ROM for the trajectories specified by $\px$ when only a rough knowledge of latter density is available but some partial observations of the trajectories are available? %More specifically, we will assume that the ROM builder has only access to the following sources of information:\\
More specifically, we will assume that we have  the following two sources of information at our disposal in the ROM construction process:\\
\begin{itemize}
\item \textit{a surrogate density} $\spx$: %\remCH{le terme ``surrogate density'' ou ``surrogate model'' ne serait-il pas plus appropri\'e? Je pense que le terme ``prior'' devient un obsolete dans la nouvelle formulation du problem car $\px$ est plus une "distribution cible" qu'un "prior"}
 this density gathers all the information the practitioners may have about the system of interest. {This density is very general in the sense it can be of any form and it does not need to satisfy any particular  constraints. } For example, one may know that the trajectories of interest obey \eqref{eq:model_init} for some  parameters included in the set $\sparamSet$. However, the true parameter set $\paramSet$ and the distribution of $\param$ over this set may be unknown. In this case, the surrogate density $\spx$ could for example be defined via \eqref{eq:model_init} by using a uniform distribution on $\param$ over $\sparamSet$. %\remCH{Dire qu'on impose pas de contraintes particulieres sur  $\spx$?}
\item \textit{incomplete observations on the target trajectories}: we assumed that ``incomplete'' observations of the trajectories are available; these observations, say $\obs$, are supposed to obey a \textit{known} conditional model $\pyx$. The term ``incomplete'' refers to the fact a realisation  $\x$ of $\px$ cannot be unequivocally recovered from its observation $\y$ by inverting the observation model. This situation  occurs for instance when only a subset of components of $\x$ are observed or when the observations are corrupted by some noise. 
 As an applicative example,  in geophysics, meteorological sensors  only provide low-resolution  and noisy observations of the atmosphere state. 
% that we observe  (\ie a realisation  $\x$ of $\px$ cannot be unequivocally recovered  by inverting the observation model). %Moreover, these observations may possibly be corrupted by noise. 
% As an example,  in geophysics, meteorological sensors  provide low-resolution  and noisy observations of the atmosphere state, which are insufficient to recover the distribution of temperature at fine scales.  \\
\end{itemize}

The main goal of this paper is therefore to propose a methodology taking benefit from these two sources of information to build  a ``good'' ROM for trajectories distributed according to $p_\X$. 

%\subsection{Problematic}
%We are now ready to expose the problematic of this paper: how to take benefit from these {incomplete} observations and  the surrogate density  to  build a  ``good'' ROM for trajectories distributed according to $p_\X$?
Before describing the contributions of this paper, we provide an overview of some state-of-the-art methodologies dealing with the problem of ROM construction from incomplete observations.   
The first contribution  dealing with this type of problem is the ``Gappy POD'' technique proposed by Everson and Sirovich in \cite{Everson1995KarhunenLoeve}. {The authors propose to  construct an approximation subspace for trajectories distributed according to $p_\X$ relying  on  the observed components of $\x$.} %\remCH{Decrire la methode en deux mot avant de dire ses defauts} 
However, this method releases poor ROM approximations  as soon as some directions of the space embedding  the  trajectories of interest  are never observed \cite{Gunes2006Gappy}. This is for example the case when these trajectories are incompletely observed through the \textit{same} observation model.

In order to circumvent this issue,  recent works  combine an observation model with  a surrogate density, in the case of the reduction of a \textit{static} high-dimensional system. %The term ``static'' refers here to the fact that the systems do not involve any temporal dimension, by opposition with the {\it dynamical} system of the form  \eqref{eq:model_init} studied in this paper \remCH{NB: je ne suis pas sur que l'aspect dynamique ou statique change grand chose au probleme...}. 
On the one hand, several authors propose this observation and prior knowledge combination in a   noise-free  deterministic  setting.~In \cite{NME:NME4747}, the authors suggest  to iteratively enrich the ROM by using point-wise estimates obtained from linear observations and a surrogate  model. In \cite{HerzetHeasDremeau16}, the authors propose to refine this approach by including the uncertainty inherent to the point-wise estimates in the reduction process.  Stable recovery guarantees are also provided from a worst-case perspective. %These two studies concentrate on a noise-free  deterministic setting.
On the other hand,  several works have investigated the context  of combining noisy observations with a  probabilistic prior.  The methodologies naturally rely in this case on  posterior probabilities \cite{CuiEtAl2014,CuiMarzoukWillcox2014,SpantiniEtAl2015}. More precisely,  in \cite{CuiMarzoukWillcox2014}, the author feed a reduced-basis technique  with samples of the posterior. In \cite{CuiEtAl2014,SpantiniEtAl2015}, an optimal  low-dimensional subspace projection of the posterior distribution is inferred based on its local Gaussian structure. 

In this paper, we  propose a general data-driven methodology for the reduction of parametric  dynamical systems, exploiting incomplete observations.  The proposed procedure exploits the two sources of information mentioned previously, namely:  
 \textit{(i)} a surrogate probabilistic characterisation %\remCH{j'ai enleve ``of the uncertainty"; je pense que l'incertitude n'est plus vraiment d'a propos ici. On a juste un modele approximatif pour $\px$.} 
 of the trajectories of interest,
\textit{(ii)} incomplete observations of these trajectories. The proposed ROM construction  relies on the minimisation of the  expectation of a bound  on the error  between the true and reduced trajectories. The expectation relies on a new data-enhanced surrogate density, say $\hat p_\X$, inferred from the initial surrogate $\spx$ and the partial observations. 
% \remCH{ce n'est plus vraiment une posterior expectation: en fait, on calcule une nouvelle densit\'e surrogate, disons $\ppx$, en exploitant les observations. Les probas a posteriori n'apparaissent que parce qu'on effectue une approximation MC sur $p(\y)$. Il faut donc reecrire cette partie je pense.} 
  An approximated solution to this minimisation problem is efficiently computed using Monte-Carlo (MC) and Sequential Monte-Carlo (SMC) techniques. The proposed approach relies on the following assumptions: 
\begin{itemize}
%approximation of the posterior expectation.
\item   stability of ROM inference when using the surrogate  $\tilde p_\X$ in place of  $ p_\X$,
\item   tightness of the  error  bound,
\item   accuracy of  the expectation approximation by MC and SMC techniques.%, such as sequential importance sampling or particle filtering, used for the 
\end{itemize}  
These properties are discussed  and  empirically assessed in the context of our numerical simulations.
 The present work complements and generalises the works  \cite{NME:NME4747,HerzetHeasDremeau16,CuiEtAl2014,CuiMarzoukWillcox2014,SpantiniEtAl2015} in two main respects: it proposes a methodology extending these works to the case of dynamical systems;  it provides a Bayesian framework  generalising  any standard ROM construction to the setup where trajectories to be reduced are not fully known.
%It focus in particular on POD-Galerkin projections and linear approximations based on DMD.   
  
 % We focus on the family of ROMs based on POD-Galerkin projection or on a noisy version of Koopman expansion  known as principal oscillating patterns (POP). We recast these ROM problems  within a probabilistic framework, derive their closed-form solutions and propose efficient sampling algorithms to  approximate them.
%We illustrate the proposed approach by the reduction of a system of Rayleigh-B\'enard convection, using partial and noisy observations. 
%\remPH{... a finir} 

%\subsection{Paper Organisation and Notations}
The rest of this paper is organised as follows. {Section~\ref{sec:target} first introduces  the target problem and  presents its surrogate analog. Section~\ref{sec:implementation} then discusses implementation issues and the MC simulation techniques used to obtain a tractable method.}~Section~\ref{sec:ex} continues by detailing the particularisation of this methodology to the context of  Galerkin projections or low-rank linear approximations. The ability of the method to take into account uncertainty  is  discussed at the end of this section.~A numerical evaluation of the proposed methodology is exposed in Section~\ref{ex:Rayleigh} and  conclusions are finally drawn in a last section.

 %\remPH{J'ai reporté la SVD et la pseudo inverse au seul endroit ou on l utilise (dans la section DMD)} 
 We will use %\remCH{employ = "engager"; attention, tu anglicises souvent des mots sans verifier au dictionnaire} 
in what follows some notations. Random vectors will be denoted by uppercase letters (as $X$) and their realisations by lowercase letters (as $x$). Boldface letters (as $\mathbf{x}$) will indicate matrices, and will be uppercase (as $\mathbf{X}$) for random matrices. $p_\X$ will refer to the probability density of  $\X$. {When there is no ambiguity, the density subscript  will be omitted to lighten notations, \ie $\px(\x)=p(\x)$.}
The symbol $\|\cdot\|_F$ and $\cdot^\intercal$ will respectively refer to the Frobenius norm and the transpose operator; %\remCH{rajouter la SVD et la definition de la pseudo inverse via la SVD}
$\mathbf{i}_k$ will denote the $k$-dimensional identity matrix.
%For any integrable function $\varphi$, expectation with respect to the probability density  $p_\X$ of $\X$ \remCH{Il serait plus logique d'ecrire $\px$ etant donne que $\X$ represente la variable aleatoire. J'ai utilise cette convention dans mes macros que j'ai utilisees dans mes modifs. De meme je pense que l'on pourrais dire que lorsqu'il n'y pas d'ambiguite, on laisse tomber l'indice de la variable aleatoire \ie $\px(\x)=p(\x)$, $\pyx(\y\vert\x)=p(\y\vert\x)$. Ca allegerait vachement les notations. } will be denoted as
%\begin{align*}
% \langle p_\X,\varphi \rangle = \int p_\X(d\x) \varphi(\x).
%\end{align*}
%\remCH{Je pense que la notation $\int p(\x) \varphi(\x) d\x$ serait plus simple et uniformiserait les choses dans la suite ou j'ai encore trouv\'e quelques problemes de notations.}
{The definition of the Kullback-Leibler distance between two densities $p_\X$ and $\tilde p_\X$  is
\begin{align*}
\textrm{KL}(p_\X, \tilde p_\X)= \int p(\state) \log \frac{ p(\state)}{\tilde p(\state)}d\state.
\end{align*}
}
%\remCH{Rajouter definition de la distance de KL}

 %\remCH{Introduire la definition de la transpose? Dans la suite, tu fais implicitement l'hypothÃ¨se que $m\leq n$; il faudrait le dire a ce stade.}
% Consecutive elements of a sequence of vectors $\{x_{t}\}_{t=t_1}^{t_2}$    gathered in a matrix will be denoted by  $\mathbf{x}_{t_1:t_2}~= ~(x_{t_1} \cdots x_{t_2})$.
\section{Target and Surrogate Problems}\label{sec:target}

In this section, we describe the main elements characterising our ROM construction problem. We first define the performance criterion that the ROM should ideally optimise  when the target density $\px$ is known. We then discuss how to modify this target problem when only a surrogate density $\spx$ is known but some incomplete observations $\obs$ of the realisations of $\px$ are available.  

The model-order reduction problem can essentially be formulated as follows: for any choice of $\param\in\paramSet$, find an (easily-computable) approximation $\statea$ of $\state$, where $\state$ is specified by \eqref{eq:model_init}.
Most ROM techniques for dynamical models encountered in the literature impose that $\statea$ obey a recursion of the form: %\remCH{pourquoi $\tilde g_1(\theta)$ ne depend pas de $\D$? Je rajoute la dependence. De plus dans la suite tu utilises $\tilde g$. Je modifie}
\begin{align}\label{eq:model_red} 
 \left\{\begin{aligned}
& \tilde x_{t}= \tilde f_t(\tilde x_{t-1}, \theta, \D) , \\%\mbox{\quad$\forall\, s\in\Sc,t\in\Tc$},
&\tilde x_1=\tilde g(\theta,\D),
\end{aligned}\right. 
\end{align} 
where $ \tilde f_t :\ \Rr^n \times \paramSet \times \mathcal{U} \to \Rr^n$ and $\tilde g:\paramSet  \times \mathcal{U}\to \Rr^n$ are some functions specifying the ROM via the choice of parameters $\mathbf{u}\in \mathcal{U}$. The nature of $\tilde f_t$, $\tilde g$ and $\D$ depends on the family of ROMs one considers. We give two examples of choices for $\tilde f_t$, $\tilde g$ and $\D$ in Sections \ref{sec:POD}  and \ref{sec:POP}. For now, the only ingredient the reader should keep in mind is that, given a family of reduced models, the ROM is fully characterised by the choice of the parameters $\mathbf{u}\in \mathcal{U}$. 

In this respect, we will assume hereafter that an ideal choice for $\D$ is given by 
%\begin{align}\label{eq:targetProblem}
%\D^\star =\argmin_{\D \in \mathcal{U}} \Biggl\{\int  p(\state) \| \x-  \tilde \x( \textcolor{red}{\theta,}\D)  \|_F^2\, d\state\Biggr\} \quad \textcolor{red}{ \mbox{s.t.  $(\x,\theta)$  satisfy \eqref{eq:model_init}}},
%\end{align} 
\begin{align}\label{eq:targetProblem}
\D^\star =\argmin_{\D \in \mathcal{U}} \Biggl\{\int  p(\state) \| \x-  \tilde \x(\D)  \|_F^2\, d\state\Biggr\},
\end{align} 
that is, the choice of the ROM parameters should be such that they minimise the mean square approximation error over the target density $\px$. {Here, the notation $\tilde \x(\mathbf{u})$ refers to the fact $\tilde \x$ is a function of $\mathbf{u}$. Note that 
 it is also a function of parameter $\x$,  since $\tilde \x$ depends on $\theta$ which is  itself related to $\x$ through the constraint  \eqref{eq:model_init}.}
%\remCH{Mentionner que la $\tilde \x(\theta,\mathbf{u})$ notation refere au fait que $\tilde \x$ est une fonction de $\mathbf{u}$? Dans la formulation actuelle \eqref{eq:model_red}, il n'est (toujours) pas clair que $\tilde \state$ depend de $\state$.} 
Unfortunately, when $\px$ is unknown, evaluating $\D^\star$ according to \eqref{eq:targetProblem} is  not possible. One possible option to solve this problem may be to  substitute $\px$ in \eqref{eq:targetProblem} by its surrogate density $\spx$, that is
\begin{align}\label{eq:surrogateProblemdatablind}
\D^\star =\argmin_{\D \in \mathcal{U}} \Biggl\{\int  \tilde p(\state) \| \x -  \tilde \x(\D)  \|_F^2\, d\state\Biggr\}.
\end{align} 
%\textcolor{red}{ Here, on the contrary to problem \eqref{eq:targetProblem},  the constraint  \eqref{eq:model_init} has been   relaxed. Indeed, since we chose to make no particular assumptions on the surrogate density structure, there does not necessarily exist a  dependence between the realisation $\state$ and the parameter~$\theta$. In this context, recalling  explicitly that $\tilde \x$ is a function of parameter $\textcolor{red}{\theta}$ becomes obsolete so that we omit this dependence  in \eqref{eq:surrogateProblemdatablind} and in what follows we will simply denote  the approximated trajectory by $\tilde \x(\D)$.}
This formulation does however not take into account the possible presence of partial observations $\obs$ of the realisations $\state$ following  $\px$. 
In this paper, we thus propose the following alternative surrogate problem:
\begin{align}\label{eq:surrogateProblemdata}
\D^\star =\argmin_{\D \in \mathcal{U}} \Biggl\{\int  \pp(\state) \| \x -  \tilde \x(\D)  \|_F^2\, d\state\Biggr\},
\end{align} 
where $\ppx$ is defined as
\begin{align}\label{eq:defppx}
\pp(\state) = \int \pp(\state\vert\obs) p(\obs)\, d\obs,
\end{align}
with
\begin{align}\label{eq:defppxy}
\pp(\state\vert\obs) = \frac{p(\obs\vert\state) \sp(\state)}{\int p(\obs\vert\state') \sp(\state')\, d\state'}. 
\end{align}

We note that $\ppx$ obeys the standard relationship between a joint density and its marginal.~More specifically, we have from elementary probability theory that $p(\state) = \int p(\state,\obs) \,d\obs= \int p(\state\vert\obs) p(\obs)\, d\obs$. Since $\pxy$ depends on $\px$ and is therefore unknown, we propose to substitute this quantity by the surrogate posterior $\ppxy$ defined in \eqref{eq:defppxy}. Similarly, the latter surrogate verifies the standard definition of the posterior probability $\pxy$ with the difference that the target prior $\px$ has been replaced by $\spx$. 

On top of these intuitive arguments motivating the definition of $\ppx$, the following result provide a theoretical justification to \eqref{eq:defppx}-\eqref{eq:defppxy}:

\begin{proposition}\label{prop:justification_ppx}
Let $\ppx$ be defined as in \eqref{eq:defppx}-\eqref{eq:defppxy}. Then we have\footnote{The proof of the result stated in Proposition \ref{prop:justification_ppx} requires some additional technical assumptions. In order to keep the  result stated in this proposition  as simple as possible to  the practitioner, we mention these assumptions in this footnote. {The proposition holds as long as $\ppy$ satisfies
\begin{align*}
\textrm{supp}(\pyx \py) \stackrel{(e)}{\subseteq}  \textrm{supp}(\ppy) \stackrel{(f)}{\subseteq} \textrm{supp}(\py),\quad \forall \x.
\end{align*}
In particular, if for all $\x$ the density $\pyx$   has an infinite support, then these inclusions are guaranteed by the definition of $\py$ and $\ppy$. This sufficient condition is satisfied for example in the case where $\pyx$  is a model with Gaussian additive noise. Let us detail the necessity of $(e)$ and $(f)$. Inclusion  $(e)$ is needed for the  existence of the integral $\int p(\y\vert \x) \frac{p(\y)}{\pp(\y)}d\y$ and ${  \int  p(\y\vert \x) \log \frac{p(\y)}{\pp(\y)}}$, which we have assumed  to obtain ${\it (b)}$ and ${\it (c)}$.  On the other hand, to obtain ${\it (c)}$ we have applied the Jensen's inequality with the strictly convex function  $-\textrm{log}$ on the interval of  strictly positive reals. Thus, we need to check that $ \forall \y$ we have $\frac{p(\y)}{\pp(\y)}> 0$ and in particular that $\frac{p(\y)}{\pp(\y)}\neq 0$, which is guaranteed by inclusion $(f)$.%On the one hand, we can show that $\textrm{supp}(\px) \subseteq \textrm{supp}(\spx) \Rightarrow \textrm{supp}(\py) \subseteq \textrm{supp}(\ppy)$ and  that $\textrm{supp}(\spx) \subseteq \textrm{supp}(\px)\Rightarrow \textrm{supp}(\ppy) \subseteq \textrm{supp}(\py)$. On the other hand, we obviously have $\textrm{supp}(\pyx \py) \subseteq \textrm{supp}(\py)$ for any $\x$. This show that   $e)$ and  $f)$ hold if  $\textrm{supp}(\spx) = \textrm{supp}(\px)$.  
} }\vspace{0.2cm}
\begin{align}
\mathrm{KL}(\px,\ppx) \leq& \mathrm{KL}(\px,\spx) - \mathrm{KL}(\py,\ppy),
\end{align}
where
\begin{align}
\pp(\obs) = \int p(\obs\vert\state) \sp(\state)\, d\state. \\
\nonumber
\end{align}
\end{proposition}
\textit{Proof}: The result is a consequence of the following inequalities:
\begin{align}
\mathrm{KL}(p(\x),p^{(k)}(\x))
%&= \int_\x p(\x) \log \frac{p(\x)}{p^{(k)}(\x)}\nonumber\\
&\stackrel{(a)}{=} \int  p(\state) \Bigl(\log p(\state) -\log \pp(\state)\Bigr)d\state,\nonumber\\
&\stackrel{(b)}{=} \int  p(\state) \Bigl(\log p(\state)-\log \sp(\state)-  \log \int p(\y\vert \x) \frac{p(\y)}{\pp(\y)}d\y \Bigr)d\state,\nonumber\\
&\stackrel{(c)}{\leq} \int  p(\state) \Bigl(\log p(\state)-\log \sp(\state)-  {  \int  p(\y\vert \x) \log \frac{p(\y)}{\pp(\y)}}d\y \Bigr)d\state,\nonumber\\
&\stackrel{(d)}{=}  \int  p(\x) \Bigl(\log p(\x)-\log \sp(\x)\Bigr)d\state - \int  p(\y) \Bigl(\log p(\y)- \log \pp(\y)\Bigr)d\y,\nonumber
\end{align}
where $(a)$ follows from the definition of the Kullback-Leibler distance, $(b)$ from the definition of $\ppx$ in \eqref{eq:defppx}-\eqref{eq:defppxy}, $(c)$ is a consequence of the Jensen's inequality, and $(d)$ follows from $\int p(\obs\vert\state) p(\state) \,d\state = p(\obs)$. $\square$\\
 
%\addCH{Peaufiner la preuve. Discuter les conditions qui permettent d'appliquer l'inegalite de Jensen. }\\

The operational meaning of Proposition \ref{prop:justification_ppx} is as follows: as far as the Kullback-Leibler distance is considered, the approximation of $\px$ by $\ppx$ is at least as good as the approximation of $\px$ by our  surrogate $\spx$. Moreover, when $\mathrm{KL}(\py,\ppy)>0$, the proposed approximation $\ppx$ leads to a strict improvement of the initial surrogate $\spx$, that is $\mathrm{KL}(\px,\ppx) < \mathrm{KL}(\px,\spx)$. 
 On the one hand, $\ppy$ can be understood as the distribution that the observations should obey if the state variable $\state$ was distributed according to $\ppx$.~On the other hand, $\py$ corresponds to the actual distribution of the collected observations $\y$. Since $\mathrm{KL}(\py,\ppy)=0$ if and only if $\py(\obs)=\ppy(\obs)$ (almost everywhere), we see that $\ppx$ leads to a strict improvement over $\spx$ as soon as the empirical distribution of the collected data deviates from the surrogate $\ppy$.

%We can make the following comments about the surrogate density $\ppx$ appearing in \eqref{eq:surrogateProblemdata}. 

We can notice that $\ppx$ depends on the observations $\obs$  via the distribution $\py$ in \eqref{eq:defppx}.  {The precise knowledge of $\py$ is however inaccessible in most experimental setups   and in order to build $\ppx$, the  practitioner can only access to a finite set of realisations of the observed  random variable. We will detail in the following how $\ppx$ is approximated  from this finite set of observations.}

\section{Implementation Issues}\label{sec:implementation}

 The target problem exposed in the previous section is intractable directly. Its resolution will rely on several  levels of approximations. They are detailed below.
\subsection{MC  Approximation}\label{sec:mc}

%\remCH{Il n'est pas vrai necessaire de passer par la generation des $x^(i)$. On peut dire directement que l'on a acces a des realisations $y^{(i)}$ de $\py(\obs)$, non? Du coup, je pense qu'on peut simplifier grandement cette section.}
In practice, we are often faced to the lack of knowledge of the density $\py$. Nevertheless, a practitioner may have  access to a finite set of   partial  observations   $$\{\mathbf{y}^{(i)} \in \Rr^{m\times T }; \quad i=1,\cdots, D\},$$  where $m < n$, composed of realisations assumed independent and identically distributed ({\it i.i.d.})  according to the density $\py$. 
Relying on these observations,  we propose to approximate the marginalisation integral in \eqref{eq:defppx} {via} a standard  MC technique. This leads to the following approximation of the cost function     in problem \eqref{eq:surrogateProblemdata} 
\begin{align}\label{eq:targetProblemMC}
 %&= \argmin_{\D \in \mathcal{U}} \langle \int_{\y \in \Rr^{m \times DT}}p_{\x|\mathbf{y}}(d\x,\mathbf{y}) p_{\mathbf{y}(d\mathbf{y})},  \| \x -  \tilde \x(\D)  \|\rangle , \nonumber \\
 \int \pp(\state)   \| \x -  \tilde \x(\D)  \|^2_F d\x \simeq \frac{1}{D}\sum_{i=1}^D  \int \pp(\state\vert\obs^{(i)})   \| \x -  \tilde \x(\D)  \|^2_F d\x .
\end{align} 
%\remCH{Au lieu de \eqref{eq:targetProblemMC}, je mettrais plutot (cela met plus en evidence que l'approximation prote au niveau de la fonction cout)}:
%\begin{align}\label{eq:targetProblemMC2}
%\int  \ppx(\state) \| \x -  \tilde \x(\D)  \|_F^2\, d\state \simeq \sum_{i=1}^D   \int p_{\x|\mathbf{y}^{(i)}}(\state)\, \| \x -  \tilde \x(\D)  \|^2_F \, d\state
%\end{align}
{We remark that, by introducing this MC approximation, observations $\obs^{(i)}$ now appear explicitly in the ROM inference problem, on the contrary to the cost function in problem \eqref{eq:surrogateProblemdata} which only exhibits a dependence to the unknown density $\py$.
%\remCH{Mentionner que les observations $\obs^{(i)}$ interviennent maintenant directement dans la fonction cout (contrairement a ce que l'on avait dans \eqref{eq:surrogateProblemdata} ou $\ppx$ dependant juste de $\py$} 
%For large $D$, the solution of \eqref{eq:targetProblemMC} is an accurate approximation of $\D^\star$ \remCH{Ton statement est un peu trop fort a mon avis: }
It is well know that the right-hand side of equation \eqref{eq:targetProblemMC} is an unbiased estimate of the left-hand side with an error variance evolving as $\mathcal{O}(D^{-1})$}. 
%Indeed, it is well know that the arithmetic average in \eqref{eq:targetProblemMC} will yield an unbiased estimate of the expectation with respect to $p_\mathbf{y}$ and the variance of the estimation error will converge to zero  at the standard rate of $1/{D}$.
    
 \subsection{SMC Approximation}\label{sec:smc}
 
  {
     In the general case, the   density $\pp_{\x|\mathbf{y}^{(i)}}$ appearing in \eqref{eq:targetProblemMC} is not closed-form and we can often not   compute analytically posterior expectations. % with respect to the posterior. %, such as  the objective in \eqref{eq:romDeter}. %since in most cases a closed-form expression is rarely available for probability measure $\nu$ or $\bar \nu$. 
     We pursue an approximation of this  density   for $i=1,...,D$ by an empirical  measure  of the form %\remCH{je change $\x\rightarrow\X$, $\y\rightarrow\Y$, $d\x\rightarrow\x$ ci-dessous}
 \begin{align}\label{eq:romCrit2}
&\pp(\x\vert\obs^{(i)}) \simeq \frac{1}{N} \sum_{j=1}^{N} w^{(i,j)} \delta_{\XI^{(i,j)}}(\x),
\end{align} 
which relies on a set of $N$ samples  $\{\XI^{(i,j)}\}_{j=1}^{N}$, weights $\{w^{(i,j)}\}_{j=1}^{N}$ with $\XI^{(i,j)}=(\xi^{(i,j)}_1\cdots\xi^{(i,j)}_T)\in \Rr^{n \times T}$ and $w^{(i,j)} \in \Rr_+$ and  the Dirac measure $\delta_{\x}$.  This leads to an approximation of  the cost function     in problem \eqref{eq:surrogateProblemdata}  by the weighted sum
 \begin{align}\label{eq:romMC}
 \int \pp(\x|\mathbf{y}^{(i)}) \| \x -  \tilde \x(\D)  \|_F^2d\x  \approx \frac{1}{N} \sum_{j=1}^{N} w^{(i,j)}  \| {\XI}^{(i,j)} -  \tilde \x(\D)  \|^2_F,    
\end{align}
% where $ \tilde \x(\D)$ depends on the initial state of the particle ${\XI}^{(i,j)}$.
and combining  approximations   \eqref{eq:targetProblemMC} and \eqref{eq:romMC}, we obtain
%\begin{align}\label{eq:targetProblemMCSMC}
% %&= \argmin_{\D \in \mathcal{U}} \langle \int_{\y \in \Rr^{m \times DT}}p_{\x|\mathbf{y}}(d\x,\mathbf{y}) p_{\mathbf{y}(d\mathbf{y})},  \| \x -  \tilde \x(\D)  \|\rangle , \nonumber \\
%& \argmin_{\D \in \mathcal{U}} \frac{1}{DN}\sum_{i=1,j=1}^{D,N}  w^{(i,j)}  \| {\XI}^{(i,j)} -  \tilde \x(\D)  \|^2_F.
%\end{align} 
%
%\remCH{Ici, de la meme facon que je l'ai suggere dans la sous-section precedente, je remplacerais \eqref{eq:romMC} et \eqref{eq:targetProblemMCSMC} par}
\begin{align}\label{eq:targetProblemMCSMC}
\int  \pp(\state) \| \x -  \tilde \x(\D)  \|_F^2\, d\state \simeq \frac{1}{DN}\sum_{i=1}^{D}\sum_{j=1}^{N}  w^{(i,j)}  \| {\XI}^{(i,j)} -  \tilde \x(\D)  \|^2_F. 
\end{align}
%\remCH{Donc en definitive, juste garder \eqref{eq:romCrit2} et \eqref{eq:targetProblemMCSMC} permet de mieux cerner l'essence de l'approximation.}
% 
% \remCH{Introduire ensuite le prior surrogate particulier \eqref{eq:recOrginalNon}-\eqref{eq:recOrginalNon2} et la factorisation particuliere de $\pyx$, cad:}
% \begin{align}
%\pyx = \prod_{t=1}^T p_{Y_t\vert X_t}.
%\end{align}
%
}
{
In the case of dynamical systems, $\ppxy$ in \eqref{eq:romCrit2} often exhibits a nested structure which can be sampled in a sequential manner.   
In particular, the surrogate density $\spx$, used for defining $\ppxy$ in \eqref{eq:defppxy}, often takes the form of a Markov chain defined  by a
  transition  kernel and an initial density
\begin{align}\label{eq:recOrginalNon}
 \left\{\begin{aligned}
 \tilde p(x_t| x_{t-1})&=\pi_t(x_t,x_{t-1}),\\%\mbox{\quad$\forall\, s\in\Sc,t\in\Tc$},
\tilde p(x_1)&=\eta_1(x_1),
\end{aligned}\right. \vspace{-0.cm}%\\
\end{align} 
which will imply the   density factorisation
%\begin{align*}
%&
\begin{align}
\tilde p(\x)= \eta_1(x_1)\prod_{t=2}^T \pi_t(x_t,x_{t-1}).\label{eq:recOrginalNon2}
\end{align}
}

{
SMC techniques are particularly well suited to this context  and constitute tractable methods able to  compute efficiently a relevant set of  $N$ samples and weights  $\{\XI^{(i,j)},w^{(i,j)}\}_{j=1}^{N}$   involved in  \eqref{eq:romCrit2}. Among the  variety of SMC techniques, the most known methods are sequential importance sampling or  bootstrap particle filtering~\cite{doucet2000sequential}. These algorithms exploit an observation model admitting  the  factorisation 
 \begin{align}\label{eq:factorYX}
\pyx = \prod_{t=1}^T p_{Y_t\vert X_t},
\end{align}
%
%
%
%an observation model of the form  \vspace{-0.cm}
%  \begin{align}
%Y_t = h_t( x_t)+W, \quad     W \sim p_{\textrm{w}}\label{eq:recYOrginalNon},
%\end{align}
% for some realisation $\x=(x_1 \cdots x_T)$. 
%This observation model involves   ${h}_t : \Rr^{n}\to \Rr^{m}$ and a sequence of $T$ noise realisations $\textrm{w}\in \Rr^m$  independent  and identically distributed according to $p_w$.  
where the random matrix  $\mathbf{{Y}} = (Y_1 \cdots Y_T)$  gathers the observed variables at the $T$ different temporal indexes.}
For large $N$,  approximation \eqref{eq:romMC} by SMC techniques is  accurate in the sense that it will yield an unbiased (or asymptotically unbiased) estimation of the posterior expectation in the case the cost function is a bounded  function of $\x$. Moreover, under this boundedness  hypothesis, the  variance of the estimation error will decrease at the rate  of $\mathcal{O}(N^{-1})$, see \eg \cite{crisan2002survey}.  However, in the our case, the error norm $ \| \x -  \tilde \x(\D)  \|^2_F$ is in general not  bounded. Although progress has  been recently accomplished in this direction~\cite{2015arXiv151106196A},  extending these asymptotical unbiased properties and convergence results  to the case of unbounded test functions has not been done yet in the context of SMC approximations.%,HeasLegland??}.\\

\subsection{Practical Identification of a Minimiser} 

%\remCH{Je te propose une formulation alternative a cette section. Ma motivation est de faire disparaitre la dependence du probleme d'optimisation que tu obtiens dans le sous-espace $\mathcal{S}$ (qui n'a pas ete defini auparavant et la distance $\mathrm{dist}$. En fait, je pense qu'on peut dire les choses plus simplement en mentionnant juste qu'on optimise une borne sup (plus ou moins tight) sur la fonction cout de notre probleme initial (la notion de sous-espace et de distance n'est pas necessaire ici). Les exemples de borne sup dans lesquels tu fais apparaitre des sous-espaces $\mathcal{S}$ peuvent etre deplace a la section suivante dans les exemples. Je mentionne egalement a la fin que, bien qu'on considere les approches de type "upper bound" dans la suite, rien ne nous empeche en principe de considerer les methodologies de type "optimisation locale". }

{With the simplifications proposed in Sections \ref{sec:mc} and \ref{sec:smc}, our surrogate optimisation problem takes the form:
\begin{align}\label{eq:surrogateOP}
\D^\star &= \argmin_{\D \in \mathcal{U}}  \Biggl\{ \sum_{i=1}^{D}\sum_{j=1}^{N}  w^{(i,j)}  \| {\XI}^{(i,j)} -  \tilde \x(\D)  \|^2_F \Biggr\}. 
\end{align}
%In this section, we discuss practical procedures to identify the solution of this problem. First note that 
Unfortunately, \eqref{eq:surrogateOP} is typically\footnote{That is for most choices of functions $\tilde f_t$ and $\tilde g$ encountered in practice. See also Sections \ref{sec:POD}  and \ref{sec:POP}.} a non-convex optimisation problem. Hence, designing polynomial-time optimisation procedures ensuring the identification of a global minimiser $\D^\star$ of \eqref{eq:surrogateOP} for any problem instance is usually out of reach. In order to circumvent this issue, two different approaches are usually suggested in the literature: \textit{(i)} resorting to local optimisation procedures;  \textit{(ii)} optimising an upper bound of the cost function in \eqref{eq:surrogateOP}. }

%Generally,   the target problem \eqref{eq:targetProblem} so as  problem \eqref{eq:targetProblemMCSMC}, are non-convex and intractable minimisations due in particular  to the sequential structure of \eqref{eq:model_red}. 

{The local optimisation procedures encountered in practice usually derive from iterative gradient descent methods. When the ROM approximation $\tilde{\state}(\D)$ satisfies a recursion as \eqref{eq:model_red}, these methods can be efficiently implemented by using  \textit{adjoint} procedures, see for example \cite{Hasselmann88,Kwasniok96}. The drawback of local optimisation procedures is however that they are prone to converge to local optimum of the cost function. In many situations, this behavior may prevent these methods from delivering a solution close to the global minimiser $\D^\star$, leading in turn to poor reduction performance.}

{In order to circumvent this problem, another approach pursued in the literature consists in optimising an upper bound on the cost function, that is
\begin{align}\label{eq:minupperbound}
\D^\star &= \argmin_{\D \in \mathcal{U}} \{ J(\D) \},
\end{align}
where $J(\D)$ is such that 
\begin{align}
\sum_{i=1}^{D}\sum_{j=1}^{N}  w^{(i,j)}  \| {\XI}^{(i,j)} -  \tilde \x(\D)  \|^2_F \leq J(\D)\quad \forall \D\in\mathcal{U}. 
\end{align}
Of course, the minimisers of \eqref{eq:minupperbound} usually differ from those of \eqref{eq:surrogateOP}. Nevertheless, if the behavior of $J(\D)$ is ``not too far'' from the one of $\sum_{i=1}^{D}\sum_{j=1}^{N}  w^{(i,j)}  \| {\XI}^{(i,j)} -  \tilde \x(\D)  \|^2_F$, one may expect the minimisers of $\eqref{eq:minupperbound}$ to be good approximations of the solutions of \eqref{eq:surrogateOP}. Moreover, the numerical optimisation \eqref{eq:minupperbound} may be far easier than the one of the initial optimisation problem \eqref{eq:surrogateOP}. We will provide two instances of such scenarios in Section \ref{sec:ex}.  
}

{In the sequel, we will exclusively focus our attention on methodologies based on the optimisation of an upper bound. This is motivated by the fact that, by using this approach, several methodologies known in the reduced-model community can be revisited and extended in the probabilistic framework considered in this paper. Nevertheless, the methodologies based on local optimisation procedure could be applied in a similar way in the framework discussed in this paper. }

\section{Two Examples} \label{sec:ex}
{In this section, we illustrate how the procedure presented in Section \ref{sec:implementation} particularises to two different families of ROMs. In particular, we will show that these particularisations can be seen as generalisations, in a probabilistic framework, of well-known ROM techniques, namely POD and DMD.} {We also discuss how they can be seen as generalisations of standard ROM constructions based on point estimates.}
%\remCH{Parler aussi de ce qui est fait dans la section \ref{sec:aware}. }
%We present hereafter two particularisations of our methodology  building ROMs for uncertain dynamical systems. They can be seen as a Bayesian generalisation of standard procedures for ROM construction.

\subsection{Galerkin Projection}\label{sec:POD}

The low-rank approximation called \textit{Galerkin} projection  of the dynamics \eqref{eq:model_init} 
is  obtained by   projecting  $x_t$'s onto a subspace spanned by the columns of some matrix $ \D\in \Rr^{n\times k}$ with orthonormal columns and where $k < n$, see \eg \cite{quarteroni2015reduced}. More precisely, it consists in a recursion \vspace{-0.15cm}
\begin{align}\label{eq:redModel}
 \left\{\begin{aligned}
&z_t= \D^\intercal{f}_t( \D z_{t-1}, \theta),\\
&z_1= \D^\intercal g(\theta), 
\end{aligned}\right.
\end{align} 
 \noindent defining  a sequence of $k$-dimensional variables  $\{z_t \in\Rr^k\}_{t=1}^T$. {Because $k<n$}, system \eqref{eq:redModel} is usually either tractable or efficient methods can be used to simplify computation~\cite{Barrault2004667,chaturantabut2010nonlinear}. Once recursion \eqref{eq:redModel} has been evaluated, an approximation of state $x_t$ can simply be obtained as $\tilde x_t= \D z_t$. Recursion \eqref{eq:redModel} is thus a particularisation of \eqref{eq:model_red} with 
 \begin{align*}
& \tilde f_t(\tilde x_{t-1}, \theta, \D)= \D \D^\intercal f_t( \D \D^\intercal\tilde x_{t-1},\theta),\\
& \tilde g  (\theta, \D) = \D\D^\intercal g(\theta).
\end{align*}
{Setting  $\tilde f_t$ and $\tilde g$ as above implies that  \eqref{eq:surrogateOP} with the admissible set  
 \begin{align}\label{eq:adminstePOD}
 \mathcal{U}=\{\D \in \Rr^{n\times k}| \D^\intercal\D=\mathbf{i}_k\}
 \end{align}
  is a non-convex minimisation problem exhibiting a complex sequential structure. Unfortunately, no polynomial-time optimisation methods can ensure the identification of a global minimiser in this context. %  No closed-form solution are known for this problem and one can only hope to reach a local minimum with standard descent methods.
We resort instead to the optimisation of the following upper bound of the cost function in \eqref{eq:surrogateOP}
\begin{align}\label{eq:upperBoundPOD}
J(\D)= c \sum_i\sum_j w^{(i,j)}  \| {\XI}^{(i,j)}-\D\D^\intercal {\XI}^{(i,j)} \|_F^2
\end{align}
where $c>0$.
}
%\remCH{
% Dire que cette borne est ``tight'' si $c$ est petit etant donne que 
%\begin{align}
%\| \x-\D\D^\intercal \x  \|^2_F \le \| \x-\tilde \x(\D)  \|^2_F \le c \, \| \x-\D\D^\intercal \x  \|^2_F.
%\end{align}
%Bref, enlever toutes les references a $\mathcal{S}$ et $\mathrm{dist}$ qui ne servent a present plus a rien. 
%}
%
%
{Indeed, using a generalisation of {\it C\'ea}'s lemma to  strongly monotone and Lipschitz-continuous functions~\cite{Ciarlet02}, we obtain   
%\begin{align}\label{eq:bornesError}
%\textrm{dist}^2(\x, \mathcal{S}(\D)) \le \| \x-\tilde \x(\D)  \|^2_F \le c \, \textrm{dist}^2(\x, \mathcal{S}(\D)),
%\end{align}
\begin{align}\label{eq:bornesError}
\| \x-\D\D^\intercal \x  \|^2_F \le \| \x-\tilde \x(\D)  \|^2_F \le c \, \| \x-\D\D^\intercal \x  \|^2_F.
\end{align}
for any $\x \in \Rr^{n\times T}$ and   with the constant $c$ independent of $\D$. This bound is tight for small $c$, since it encloses from above and below the error norm. 
%  in terms of its  distance 
%  \begin{align}\label{eq:distPOD}
% \textrm{dist}(\x, \mathcal{S}(\D)) = \| \x-\D\D^\intercal \x  \|_F,
%  \end{align}
%   to the approximation subspace  
% \begin{align*}
%\mathcal{S}(\D)=  \underbrace{\textrm{im}(\D ) \times \cdots \times \textrm{im}(\D )}_{T\,\, \textrm{times}},
% \left\{\begin{aligned}
%& \Rr^n\quad\quad\quad\,\,\, \textrm{for } \quad t=1,\\
%&\textrm{im}(\D \mathbf{w}_\D) \quad \textrm{for} \quad t=2,\cdots,T.  \\
%\end{aligned}\right.
%\end{align*}
The upper bound in \eqref{eq:bornesError} is obtained under the assumption that the mapping $\Rr^{n \times T}  \to \Rr^{n \times T},  \x  \to (x_1,-f_1(x_1)+x_2,-f_2(x_2)+x_3,\cdots,-f_{T-1}(x_{T-1})+x_T)$ is strongly monotone and that it is Lipschitz-continuous for bounded arguments  \cite[Theorem 5.3.4]{Ciarlet02}. 
We remark that the optimisation of the bound \eqref{eq:upperBoundPOD} can be seen as a POD-like problem, where standard snapshots are substituted by weighted samples obtained by MC and SMC simulations.
}

Problem \eqref{eq:minupperbound} with the upper bound  \eqref{eq:upperBoundPOD} and the admissible set \eqref{eq:adminstePOD}  admits a closed-form solution $\D^\star$.  Indeed, a well-known result is that the columns of matrix $\D^\star$  are the eigenvectors of  matrix $\mathbf{c}\mathbf{c}^\intercal$ where  
$$ \mathbf{c}  = (\sqrt{w^{(1,1)}} \XI^{(1,1)}\cdots \sqrt{w^{(D,N)}} \XI^{(D,N)}) \in \Rr^{n \times TDN},$$ 
associated to its $k$ largest eigenvalues \cite{jolliffe2002principal}. In practice,  eigenvectors of interest can be derived from the  eigen-decomposition of the smaller matrix $\mathbf{c}^\intercal \mathbf{c}$, see \eg \cite{quarteroni2015reduced}. 

According to the convergence results of MC and SMC techniques, we also see that  $\D^\star$  is  an unbiased estimator of the closed-formed solution of 
\begin{align}\label{eq:probOrigPOD}
&\argmin_{\D \in \mathcal{U}} \,c \int  \pp(\state) \| \x -  \D\D^\intercal \x  \|_F^2\, d\state.
\end{align}
%\ie of the non-approximated counterpart of problem    \eqref{eq:minupperbound} defined with the upper bound  \eqref{eq:upperBoundPOD}. 
We thus have the following proposition.  %the eigenvectors  of matrix $\mathbf{c} \mathbf{c} ^*$  are   unbiased estimators of the columns of  $\D^\star$, the matrix solving  \eqref{eq:romCrit3Last}.   
%( \mathbb{E} |\langle  \mu^N - \mu, \D^\star (\D^\star)^* \rangle|^2)^{1/2}  \le

\begin{proposition}\label{prop11}
 For any positive integer $k \le TDN$ the  eigenvectors corresponding to  the $k$ largest eigenvalues  of matrix $\mathbf{c} \mathbf{c} ^\intercal$  are   the columns of the solution of  \eqref{eq:minupperbound} with the upper bound  \eqref{eq:upperBoundPOD} and the admissible set \eqref{eq:adminstePOD}. Moreover, they are unbiased  estimators of the columns of the matrix solving \eqref{eq:probOrigPOD}.  
 \end{proposition}

%We highlight in Section~\ref{sec:aware} how uncertainty  impacts the computation of the eigendecomposition of $\mathbf{c} \mathbf{c} ^\intercal$, and show the limitation of inference of ROMs relying on point estimates.

\subsection{Low-Rank Linear Approximation}\label{sec:POP}

%
%
%
%Let matrix $\mathbf{g}  \in \Rr^{n \times n} $ be of rank $k$ with $k \ll n$ and where elements of the set $\{W_1, \dotsm, W_T\}$  are independent residues of probability measure $ p_{t}^{W}$, with zero mean and value in $\Rr^n$. 
%A principal oscillating patterns (POP)   approximation is
%\begin{equation}\label{eq:GaussLinSystAlter}
% \left\{\begin{aligned}
%& X_t=\mathbf{g}^{t-1}  X_{1} + W_t \quad\textrm{with}\quad W_t \sim  p_t^{W}(dw_t).\\
%&X_1 \sim \eta_1(dx_1),  \\
%\end{aligned}\right.
%\end{equation}
%where, to be equivalent to system  \eqref{eq:MarkovModel},  the probability measures $ p_{t}^{W}$  differ from the one  defined in \eqref{eq:GaussLinSyst}.
% Note that, the simple linear, stationary and low-order dynamics  $\mathbf{g}$ in \eqref{eq:GaussLinSyst} or \eqref{eq:GaussLinSystAlter} is counterbalanced by the non-linearity, non-stationarity and high-dimensionality of the sequence of  residues. The POP approximation of \eqref{eq:MarkovModel} will be a representation \eqref{eq:GaussLinSyst} or  \eqref{eq:GaussLinSystAlter}   minimising the  variance of the sequence of $W_t$'s  \cite{Hasselmann88,Kwasniok96}. 

%The POP   approximation  of the dynamics \eqref{eq:model_init}  is a Krylov subspaces approximation \cite{??} 
 A low-rank linear approximation of the dynamics \eqref{eq:model_init}  is    a particularisation of \eqref{eq:model_red} to %\remCH{formater l'equation suivante comme \eqref{eq:GaussLinSystReduced} cad avec un seul numero pour les 2 equations}
\begin{align}\label{eq:POPrecHD}
&\tilde f_t(\tilde x_{t-1}, \theta, \D)= \D\tilde x_{t-1}, \nonumber \\
& \tilde  g (\theta, \D) =  g( \theta),
\end{align}
 parameterised by  some matrix $\D\in \Rr^{n \times n}$ of rank lower or equal to $k\le n$. Let its singular value decomposition  (SVD)  be
 $
 \D=\mathbf{w}_\D\boldsymbol{\sigma}_\D\mathbf{v}_\D^\intercal,
 $
% \remCH{Les matrices doivent etre en minuscules grasses}
  with $\mathbf{w}_\D,\mathbf{v}_\D\in \Rr^{n \times k}$ and $\boldsymbol{\sigma}_\D\in \Rr^{ k \times k}$ so that $\mathbf{w}_\D^\intercal\mathbf{w}_\D=\mathbf{v}_\D^\intercal\mathbf{v}_\D=\mathbf{i}_k$ and $\boldsymbol{\sigma}_\D$ is diagonal.
%$ \D=( \R, \Q)\in \Rr^{n \times k} \times \Rr^{n \times k}$.
The  $n$-dimensional reduced  states $\{\tilde{x}_t\}_{t=1}^T$ are  fully determined  by the following recursion, %. % Indeed, we introduce the adjoint of the pseudo-inverse of  $\hat \R$  given by  $\G=\hat \R(\hat \R^\intercal\hat \R)^{-1}\in \Rr^{n \times k}$. 
%Furthermore, we choose $\mathbf{b} \in  \Rr^{k \times k}$ such that $\hat \Q= (\G)^\intercal \mathbf{b}^\intercal$, which implies that $\mathbf{b}=\hat \Q^\intercal\hat \R$.  Therefore for given matrices $\hat \R$ and $\hat \Q$, the matrice $\mathbf{b}$ is uniquely determined.   
%Using the latter matrices, the  $n$-dimensional  model \eqref{eq:POPrecHD}  is characterised  by the   k-dimensional  recursion 
\begin{equation}\label{eq:GaussLinSystReduced}
 \left\{\begin{aligned}
& z_t=  (\mathbf{v}_\D\boldsymbol{\sigma}_\D)^\intercal \mathbf{w}_\D  z_{t-1} ,\\
&z_2=(\mathbf{v}_\D\boldsymbol{\sigma}_\D)^\intercal\mathbf{w}_\D^\intercal  g(\theta), 
\end{aligned}\right.
\end{equation}
only involving $k$-dimensional variables. 
%implying the projected variable  $z_t \in \Rr^k$. %, where $\R=\mathbf{w}_\D$ and $\Q= \mathbf{v}_\D\boldsymbol{\sigma}_\D$.
 By multiplying both sides of \eqref{eq:GaussLinSystReduced} by $\mathbf{w}_\D$,   we  obtain low-rank approximations $\tilde x_{t} =  \mathbf{w}_\D  z_t$ of the $n$-dimensional states $x_t$ defined in \eqref{eq:model_init}.
%\begin{equation}\label{eq:GaussLinSystApprox}
% \left\{\begin{aligned}
%&\hat \R  z_t= \hat \R\hat \Q^\intercal\hat \R z_{t-1},\\
%&\hat \R z_1=\hat \R\G\theta_1.  \\
%\end{aligned}\right.
%\end{equation}

{
 Setting  $\tilde f_t$ and $\tilde g$ as in \eqref{eq:POPrecHD} and the admissible set  as
 \begin{align}\label{eq:adminstePOP}
 \mathcal{U}=\{\D \in \Rr^{n\times n}| \textrm{rank}(\D) \le k,  \|\D\|_{2,2} \le \lambda\}
 \end{align}
 defines  a non-convex problem  \eqref{eq:surrogateOP}  due to the low-rank constraint and the sequential structure of \eqref{eq:POPrecHD}. The  global minimiser is out of reach in this context. %  No closed-form solution are known for this problem and one can only hope to reach a local minimum with standard descent methods.
Here again, we choose to resort to the optimisation of the following upper bound of the cost function in \eqref{eq:surrogateOP}:
\begin{align}\label{eq:upperBoundPOP}
J(\D) = c \sum_{i=1,j=1}^{D,N}  w^{(i,j)}  \sum_{t=2}^T \| {\xi}_t^{(i,j)}-\D {\xi}_{t-1}^{(i,j)}  \|^2_F
\end{align}
where $c>0$  depends on  $\lambda$.
Indeed, for any $\x \in \Rr^{n \times T}$,  it is shown in  Appendix~\ref{app:2} that %, assuming there exists $\lambda\in \Rr$  such that $ \|\D\|_{2,2}\le \lambda$, 
the ROM  error 
%  \begin{align}\label{eq:appropxPOPerror}
%  \parallel \x  -\tilde \x(\D)  \parallel^2_F = \sum_{i=2}^T \parallel x_{i} -\D^{(i-1)} g( \theta) \parallel^2_2,
% \end{align} 
can be bounded for any $\D \in \mathcal{U}$ as 
\begin{align}\label{eq:boundDMD}
\sum_{t=2}^{T}\|x_{t}-\D x_{t-1}\|^2_2 \le \| \x-\tilde \x(\D)  \|^2_F \le c \,\sum_{t=2}^{T}\|x_{t}-\D x_{t-1}\|_2^2. %,
\end{align}
As in the previous example, this bound is tight as long as $c$ is small since it encloses from above and below the error norm.  

A reasonable condition  to  set $\lambda$ is that \eqref{eq:adminstePOP} includes at least   the minimisers over the unconstrained domain  
$ %\begin{align}%\label{eq:adminstePOP'}
\{\D \in \Rr^{n\times n}| \textrm{rank}(\D) \le k\}
 $ %\end{align} 
  of the  ROM error norm and of the bound  (\ie the left and right hand side of the second inequality in \eqref{eq:boundDMD}).  Appendix~\ref{app:3} shows that these minimisers have a finite norm, \ie that there exists  $\lambda < \infty$ satisfying this condition. 
%Msuch that minimisers over the unconstrained domain  
%$ %\begin{align}%\label{eq:adminstePOP'}
%\{\D \in \Rr^{n\times n}| \textrm{rank}(\D) \le k\},
% $ %\end{align}
% are in the set \eqref{eq:adminstePOP}. 
This implies that there exists $\lambda < \infty$ such that the   constraint $ \|\D\|_{2,2} \le \lambda$  is  inactive at the minima of the bound so that the constraint can be removed from the optimisation problem.  Optimising \eqref{eq:upperBoundPOP} over \eqref{eq:adminstePOP} can in consequence be seen as a low-rank DMD-like problem, where standard snapshots are substituted by weighted samples obtained by MC and SMC simulations.
 % and independent of $\D$ \remCH{attention $\lambda$ depend de $\|\D\|_{2,2}$...}. % $c=\max_{t\in\{2,...,T\}}\{\sum_{k=t}^T (1+2(k-1)) \lambda^{2(t-2)}\}$. 
}
The optimisation of this upper bound admits a closed-form solution as  shown recently in \cite{HeasHerzet16}. Indeed, defining matrices 
%let  $\mathbf{c} = \mathbf{b}\mathbf{a}^\intercal \quad \textrm{and}\quad  \mathbf{\bar c}= (\mathbf{a}\mathbf{a}^\intercal)^{\dagger}\mathbf{a}\mathbf{b}^\intercal,$  with 
$ \mathbf{a}, \mathbf{b} \in \Rr^{n \times (T-1)DN} $ as
  \begin{align*}
 \mathbf{a} &= (\sqrt{w^{(1,1)}} \XI^{(1,1)}_{1:T-1}\cdots \sqrt{w^{(D,N)}} \XI^{(D,N)}_{1:T-1}),\\ 
 \mathbf{b}&= (\sqrt{w^{(1,1)}} \XI^{(1,1)}_{2:T}\cdots \sqrt{w^{(D,N)}} \XI^{(D,N)}_{2:T}) ,
 \end{align*}
 with the notations $\XI^{(i,j)}_{\ell:m}=(\xi^{(i,j)}_\ell\cdots \xi^{(i,j)}_m),$ the optimisation problem  can  be rewritten in the synthetic form
 \begin{align}\label{eq:ProblowrankDMD}
 \D^\star \in \argmin_{\D \in \mathcal{U}} \|\mathbf{b}-\D\mathbf{a} \|^2_F,
 \end{align}
 where $\mathcal{U}$ is defined in \eqref{eq:adminstePOP} with $\lambda=\infty$.
 %We  recognise a low-rank DMD-like problem where standard snapshots are substituted by weighted samples obtained by MC and SMC simulation according to the relaxed posterior. 
This problem admits the closed-form solution  $\D^\star=\R\R^\intercal \mathbf{b}\mathbf{a}^\dagger$, 
 where the columns of $\R\in \Rr^{n \times k}$ are real orthonormal eigenvectors associated to the largest eigenvalues of matrix $\mathbf{b}\mathbf{a}^{\dagger}\mathbf{a}\mathbf{b}^\intercal$, 
and where $\mathbf{a}^\dagger=\mathbf{v}_\mathbf{a}\boldsymbol{\sigma}_\mathbf{a}^{-1}\mathbf{w}_\mathbf{a}^\intercal$ is the Moore-Penrose pseudo-inverse of $\mathbf{a}$ \cite[Theorem 3.1]{HeasHerzet16}.
  We note that this solution can  be efficiently computed by SVDs, as detailed in \cite[Algorithm~1]{HeasHerzet16}.
 
 According to the convergence results of MC and SMC techniques, we remark that  $\D^\star$  is  an unbiased estimator of the closed-formed solution\footnote{This closed-form solution can be obtained by generalising  \cite[Theorem 3.1]{HeasHerzet16} to a continuous setting. However, we  omit    details here since  it is out of the scope of the paper. 
 }
  of 
\begin{align}\label{eq:probOrigPOP}
&\argmin_{\D \in \mathcal{U}}\, c \int  \pp(\state)\sum_{t=2}^{T}\|x_{t}-\D x_{t-1}\|_2^2\, d\state.
\end{align}
%\ie of the non-approximated counterpart of problem    \eqref{eq:ProblowrankDMD}. 
This yields the following proposition.  

 \begin{proposition}\label{prop22}
For any positive integer $k \le (T-1)DN$, $\D^\star= \R\R^\intercal \mathbf{b}\mathbf{a}^\dagger$ is a solution  of \eqref{eq:ProblowrankDMD}  and an unbiased estimator of a solution of \eqref{eq:probOrigPOP}. 
  \end{proposition}
  
%We highlight in the following  section how $\D^\star$ naturally takes into account uncertainties on the trajectories to be reduced, and show the limitation of inference relying on point estimates.

%\subsection{Remarks on Uncertainty Awareness}\label{sec:aware}
\subsection{Comparison with ROM Based on Point Estimates}\label{sec:aware}

%\remCH{Adapter les notations de cette section aux nouvelles notations}
We show that the ROM parameter $\D^\star$ inferred in  Section \ref{sec:POD} or \ref{sec:POP}  differs from the parameter inferred  relying on  point estimates \cite{NME:NME4747,Sirovich87}. The latter approach consists in building the ROM from  estimates of the state, say  $\hat{\state}^{(i)}$, computed for $i=1,...,D$ by combining the received observation $\obs^{(i)}$ and the  surrogate $\spx$. {A common choice to obtain these estimates is to rely on  the minimum mean square error (MMSE) estimator, \ie $$\hat{\state}^{(i)}=\int   \x \, \pp(\x | \mathbf{y}^{(i)})  \, d\x.$$
The parameter of a ROM  based on MMSE point estimates is then obtained by solving  
\begin{align}\label{eq:surrogateOPPointEstim}
\argmin_{\D \in \mathcal{U}}  \Biggl\{ \sum_{i=1}^{D}  \| \hat{\state}^{(i)} -  \tilde \x(\D)  \|^2_F \Biggr\}. 
\end{align}
In what follows, we will refer to this particular choice of estimator when invoking ROM based on point estimates. }

{
 %or more precisely by the particle approximation of this estimator. 
Analogously to our approach, we may obtain an  unbiased (or asymptotically unbiased) approximation of the MMSE estimator using  an SMC technique
\begin{align}\label{eq:ROMPointEstim}
\hat{\state}^{(i)} \simeq \frac{1}{N}\sum_{j=1}^{N}  w^{(i,j)}  {\XI}^{(i,j)}. 
\end{align}
Comparing \eqref{eq:surrogateOP} with the optimisation problem \eqref{eq:surrogateOPPointEstim} where the $\hat{\state}^{(i)}$'s are approximated with   \eqref{eq:ROMPointEstim},  we see that our approach can be seen as a generalisation of a point estimate approach where the approximation of $\ppxy$ relies on $N$ particles rather than on a single one.
}
%\remCH{Detailler un peu ici: dire que l'idee est de calculer une estimee $\hat{\state}^{(i)}$ by combining the received observation $\obs^{(i)}$ and the  surrogate $\spx$. Faire directement le lien avec l'estimateur MMSE mentionne plus bas, en disant qu'on focalise notre attention sur ce type d'estimateur. }
%\remCH{Avant de faire le lien entre les matrices utilisees pour definir $\D$ Ã  partir de leurs les vecteurs propres, on peut mentionner une chose plus simple a partir de \eqref{eq:romCrit2}: l'utilisation des ``point estimates'' peut etre vues comme l'approximation de $\ppxy$ par une seule particule (contre $N$) dans notre cas. }

Let us further detail  the differences between the two approaches. 
%We show how {\it uncertainty} related to the MMSE estimator is ignored in ROM inference based on point estimates. \remPH{a finir ...}
%To make this fact clear, n
Note that matrix $\mathbf{c} \mathbf{c} ^\intercal$  or  matrices $ \mathbf{b}\mathbf{a}^\intercal $ and $ \mathbf{a}\mathbf{a}^\intercal $ introduced previously are  MC and SMC approximations of  matrices of the form 
\begin{align}\label{eq:crosscov}
\int  p(\mathbf{y}) d\mathbf{y} \int \pp(\x|\mathbf{y})\,\x_{1+\ell:T}\x_{1:T-\ell}^\intercal d\x,  \quad \ell \in \{0,1\},
 % \langle  \mu(\x,\mathbf{y}), \x_{1+\ell:T}\x_{1:T-\ell}^\intercal \rangle, \quad \ell \in \{0,1\}.
 \end{align}
where we have used   the notations $\x_{\ell:m}=(x_\ell\cdots x_m) \in \Rr^{n \times (m-\ell+1)}.$
In particular,  according to \eqref{eq:targetProblemMC},  matrices  \eqref{eq:crosscov} are  approximated in our methodology by a MC technique yielding
%\begin{align*}
%	 %& \frac{1}{DN}\sum_{i=1,j=1}^{D,N}  w^{(i,j)}   \textrm{dist}^2({\XI}^{(i,j)}, \mathcal{S}(\D)) 
%	  &\frac{1}{D}\sum_{i=1}^D \int  \pp(\x|\mathbf{y}^{(i)} )\,\x_{1+\ell:T}\x_{1:T-\ell}^\intercal d\x,
%		 \end{align*}
%which can be decomposed as %\remCH{il manque probablement des indices $i$ dans le premier terme ci-dessous. }
%%\begin{align*}
%%	  &  \frac{1}{D}\sum_{i=1}^D \langle \tilde p_{\x|\mathbf{y}}(d\x),\left(\x_{1+\ell:T}-\langle \tilde p_{\x|\mathbf{y}}(d\x) , \x_{1+\ell:T}\rangle \right)\left(\x_{1:T-\ell}-\langle \tilde p_{\x|\mathbf{y}}(d\x) , \x_{1:T-\ell}\rangle \right)^\intercal\rangle\\
%%	  &+  \frac{1}{D}\sum_{i=1}^D \langle  \tilde p_{\x|\mathbf{y}^{(i)}}(d\x) , \x_{1+\ell:T}\rangle \langle \tilde p_{\x|\mathbf{y}^{(i)}}(d\x) , \x_{1:T-\ell}\rangle 
%% \end{align*}
%%
	 \begin{align}\label{eq:crosscovdec}
		 \frac{1}{D(T-\ell)}\sum_{i,t=1+\ell}^{D,T} \Biggl(  &\int  \pp(\x|\mathbf{y}^{(i)} )  \left(x_{t-\ell}-  \hat x_{t-\ell}^{(i)}  \right)\left(x_{t} - \hat x_{t}^{(i)}\right)^\intercal  d\x 
	+   \hat x_{t-\ell}^{(i)}  (\hat x_{t}^{(i)})^\intercal \Biggr),
	 \end{align}
	 where  $\hat x_{t-\ell}^{(i)}$ are MMSE point estimates.
	Note that the first term inside the brackets is the cross-covariance relative to $\pp(\x|\mathbf{y}^{(i)} )  $ of  vectors $X_{t-\ell}$ and $X_{t}$, while the second term is the square of the mean of this density. %or more precisely by the particle approximation of this estimator.
	ROMs based on point estimates rely only on the mean and ignore cross-covariance terms. They  approximate  matrices \eqref{eq:crosscov} by
	\begin{align}\label{eq:snapshots}
%	 \langle  \mu(\x,\mathbf{y}),  \x_{1+
%	\ell:T}\x_{1:T-\ell}^\intercal \rangle\approx%\sum_{t=1}^{T} \langle \nu,  x_{t}x^\intercal_{t} \rangle=
%	  \frac{1}{D(T-\ell)}\sum_{i=1,t=1+\ell}^{D,T} \langle  \mu(\x,\,{y}^i), {x}_{t-\ell}(\y^i) \rangle  \langle  \mu(\x,\,{y}^i), {x}_{t}^\intercal(\y^i) \rangle.
	   \frac{1}{D(T-\ell)}\sum_{i,t=1+\ell}^{D,T}  \hat x_{t-\ell}^{(i)}  (\hat x_{t}^{(i)})^\intercal.
	 \end{align}
Choosing  approximate  matrix  \eqref{eq:snapshots} instead of \eqref{eq:crosscovdec}  may imply  a poor approximation of \eqref{eq:crosscov}.  In particular,     \eqref{eq:crosscovdec} and \eqref{eq:snapshots} will significantly differ    in the case of large cross-covariances.   %the first term in the   brackets of \eqref{eq:crosscovdec} dominates,  matrix \eqref{eq:snapshots} will significantly differ from \eqref{eq:crosscovdec} and 
 %On the contrary, cross-covariance terms will systematically   be taken into account in \eqref{eq:crosscovdec} and the approximation will converge asymptotically to  \eqref{eq:crosscov}.
   Making a correspondence between  cross-covariances and uncertainty, this suggests that the proposed method integrates uncertainty relative to point estimates in the ROM inference process.% \ie  which may potentially impact dramatically the eigenvectors  associated to the largest eigenvalues of \eqref{eq:crosscov}. \\

\section{Numerical Evaluation}\label{ex:Rayleigh}

{We assess the proposed methodology with a standard  physical model known as Rayleigh-B\'enard convective system.  After introducing the parametric partial differential equation inducing the high-dimensional system, we provide  different variations of the  ROM building problem. They differ from each other by their underlying probabilistic models $p_\X$, $\pp_\X$ and $p_{\Y|\X}$. Based on this setup, we finally evaluate the performance of  four different sampling strategies to build a ROM in our uncertain context.}
%\remCH{Intro}

%\subsection{Rayleigh-B\'enard Convection}
\subsection{The Physical Setup}

We consider a Rayleigh-B\'enard convective system~\cite{chandrasekhar2013hydrodynamic}.  An incompressible fluid is contained in a bi-dimensional cell and is subject to periodic boundary conditions. The states of interest are the trajectories of the temperature  and  velocity fields in the cell.  

We introduce the following notations to state the evolution equations: the differential operators $\nabla=(\partial_{s_1},\partial_{s_2})^\intercal$,   $\nabla^{\perp}=(\partial_{s_2},-\partial_{s_1})^\intercal$ and $\Delta=\partial^2_{s_1}+\partial^2_{s_2}$ denote the gradient, the curl and the Laplacian with respect to the two spatial dimensions $(s_1,s_2)$; the operator $\Delta^{-1}$ is the formal representation of  the inverse of $\Delta$.  Convection is driven by the two following coupled  partial differential equations: at any  point of the unit cell $\s=(s_1,s_2)\in [0,1]^2$  and for any time  $t \ge1$, we have
\begin{align}\label{eq:RB}
 \left\{\begin{aligned}
\partial_t b(\s,t) + \mathbf{v}(\s,t) \cdot \nabla b(\s,t)-\rho\Delta b(\s,t) - \rho\nu \partial_{s_1} \tau (\s,t)&=0,\\
\partial_t  \tau(\s,t) + \mathbf{v}(\s,t) \cdot \nabla \tau(\s,t)- \Delta \tau(\s,t) -\partial_{s_1} \Delta^{-1} b(\s,t)&=0,
\end{aligned}\right. 
\end{align} 
where $\tau(\s,t)\in \Rr $ and $\mathbf{v}(\s,t) \in \Rr^2 $ are the temperature and the velocity and where the buoyancy  $b(\s,t)\in \Rr$ satisfies 
$$\mathbf{v}(\s,t)= \nabla^{\perp}\Delta^{-1} b(\s,t).$$ 
%\remCH{notation $\nabla^{\perp}$ undefined.} 
The parameters $\rho$ and $\nu$ appearing in \eqref{eq:RB} have the following physical meaning. The Rayleigh  number $\nu \in \Rr_+$  controls the balance between thermal diffusion and the tendency for a packet of fluid to rise due to the buoyancy force. The Prandtl number $\rho\in \Rr_+$ measures the relative importance of viscosity compared to thermal diffusion. These two parameters   control the coupling of the buoyancy evolution with the thermal diffusion process. In particular for $\nu=0$ and/or $\rho=0$,  the system is decoupled in the sense that the evolution of buoyancy is independent of   temperature.  

At  initial time $t=1$, the fluid in the cell is still and subject to a difference of temperature  between the bottom and the top. 
We set the  initial condition to:
\begin{align}\label{eq:initCondition}
b(\s,1)=&\pi_b\sin(a s_1)\sin(\pi s_2)+\epsilon_b(\s),\\
\tau(\s,1)=&\pi_{\tau}\cos(a s_1)\sin(\pi s_2)-\pi_{\tau'}\sin(2\pi s_2)+\epsilon_\tau(\s),\nonumber
\end{align}
 where $\pi_{b}$, $\pi_{\tau}$, $\pi_{\tau'},a \in \Rr$  are   parameters. %\remCH{quelle difference fais-tu entre une variable et un parametre?}. 
 This initial condition is equal to the solution of the Lorenz attractor~\cite{Lorenz63} up to the additive terms $\epsilon_b(\s),\epsilon_\tau(\s) \in \Rr$. %\remCH{Tu valides ma modif?}.  % by  and  in the particular case where $\nu=0$ and $\pi_b= \zeta^{-1}(\pi a)^{-2}$,  the non-linear system \eqref{eq:RB} simplifies into a linear  temperature evolution driven by 
% a  (non-stationary) buoyancy force associated to  a Taylor vortex~\cite{Taylor37}
% \begin{align}\label{eq:RBLinear}
% \left\{\begin{aligned}
%& b(\s,t)=\zeta^{-1}(\pi a_b)^{-2}{\exp^{-\zeta \pi^2a_b^2 t}} \sin(a_b s_1)\sin(\pi s_2),\\
% &\partial_t  \tau(\s,t) + \mathbf{v}(\s,t) \cdot \nabla \tau(\s,t)- \Delta \tau(\s,t) -\partial_{s_1} \Delta^{-1} b(\s,t)=0.
%\end{aligned}\right. 
%\end{align} 

We apply  a finite difference scheme on \eqref{eq:RB} to obtain  a discrete system of the form of \eqref{eq:model_init} with $x_t=\begin{pmatrix}{b}_t\\ {\tau}_t\end{pmatrix}\in \Rr^{n}$, $\begin{pmatrix}\epsilon_b\\\epsilon_\tau\end{pmatrix} \in \Rr^n$ and $n=1024$, where ${b}_t$'s, ${\tau}_t$'s, $\epsilon_b$'s and $\epsilon_\tau$'s are   spatial discretisations of respectively buoyancy and temperature fields at time $t$ and the initial condition additive terms. This discretised system constitutes the target model we want to reduce.  

\subsection{ Benchmark Problems}\label{sec:data}

{We consider different variations of the problem of ROM construction for unknown $p_\X$.  The  benchmark problems correspond to  different variations of the definition of the probabilistic models $p_\X$, $\sp_\X$ and $p_{\Y|\X}$. \\
}

{
We begin by specifying $p_\X$.  Let  $\theta_1=(a,\pi_b,\pi_{\tau},\pi_{\tau'})^\intercal\in \Rr^4$ and $\theta_2 =(\epsilon_b,\epsilon_\tau)^\intercal\in  \Rr^{1024}$  parameterise  the initial condition $x_1$ using \eqref{eq:initCondition}. Let $\theta_3=(\rho,\nu)^\intercal \in \Rr^2$ parametrise the dynamics~\eqref{eq:RB}.  We recall that we consider here a discretised version of  \eqref{eq:RB}-\eqref{eq:initCondition}   of the form of \eqref{eq:model_init}, which is parameterised by $\theta=(\theta_1,\theta_2,\theta_3)$. We specify the density $p_\X$ through the definition of  a probabilistic model for parameter $\theta$ and the use of   model \eqref{eq:model_init}. 
Note that $p_\X$ is in this configuration  a particularisation of \eqref{eq:recOrginalNon2} where the transition kernel is a Dirac measure.
We choose for $\theta$ %$\Theta=(\Theta_1,\Theta_2,\Theta_3)$ 
a  uniform distribution  on $\paramSet=  (\paramSet_1,\paramSet_2,\paramSet_3)$.
 The set  $\paramSet_1$ is chosen so that the initial condition lives at a distance  at  most of $\gamma$ from a $10$-dimensional subspace  of  $\Rr^{1024}$. The set $\paramSet_2$ is a centred  ball of $\Rr^{40}$ of radius $\gamma$. %, so that {\it around}  means at a distance  at  most of $\gamma$ from this subspace. 
   We choose   parameters  ruling the dynamics in a compact set $\paramSet_3$ in order to generate  buoyancy and  temperature evolutions in different regimes of viscosity/diffusivity and coupling/decoupling.  
 }
 
{
The surrogate density  $\tilde p_\X$ is defined in an analogous manner to $p_\X$.  The only difference with the definition of density $p_\X$ is that parameter $\theta$ %$\Theta=(\Theta_1,\Theta_2,\Theta_3)$ 
is drawn according to a surrogate  uniform distribution  on $\sparamSet=  ( \sparamSet_1,\paramSet_2,\paramSet_3)$ with  $ \sparamSet_1 \supset \paramSet_1 $. More precisely,  we fix $\sparamSet_1$ so that  the $\tilde x_1$'s live at a distance at most  of $\gamma$ from a $20$-dimensional subspace  of  $\Rr^{1024}$.
%We choose a probability density function $\eta$ corresponding  to $(\theta_1,\theta_3)$ (resp.  $\epsilon$) drawn from the uniform distribution over $(  \sparamSet_1,\paramSet_3)$ (resp. over the ball of radius $\gamma$ associated to the $\ell_\infty$ norm). 
}

{
 Let us finally specify the conditional density $\pyx$. It is chosen to be a Gaussian distribution with uncorrelated components  so that it admits  the  factorisation \eqref{eq:factorYX} where $p(y_t|x_t)$ is  a normal distribution of mean  $\mathbf{h} x_t$ with   $\mathbf{h}\in \Rr^{m \times n}$    and of covariance $ \zeta^2\mathbf{i}_{m}$  with $\zeta \in \Rr_+$.
Matrix $\mathbf{h}$ is chosen to be  a discrete approximation of the convolution by a sinus cardinal kernel so that it represents an ideal low-pass filter degrading the resolution by the factor $n/m=2$.
  }
  
 {
 Using this configuration, we are  able to generate $D$ {\it i.i.d.} realisations $\{\mathbf{x}^{(i)}, i=1,\cdots, D\}$ of $p_\X$ by uniformly sampling the set $(\paramSet_1,\paramSet_2,\paramSet_3)$ and   using model \eqref{eq:model_init}.  Drawing one sample according to each density $p(y_t|x_t^{(i)})$ then yields  the set of observations $\{\mathbf{y}^{(i)},  i=1,\cdots, D\}.$\\
 }

{
We are now ready to present the  benchmark problems.  We want to evaluate  the influence  of the following parameters: the trajectories length  $T$,  the number of observations $D$,   the noise variance $\zeta^2$,  the initial condition distribution (uniform distribution supported either on a subspace, \ie $\paramSet_2=\{0\}$, or a high-dimensional slice of thickness $2\gamma$, \ie $\paramSet_2$ is a centred ball of $\Rr^{40}$ of radius $\gamma>0$) and the set  $\paramSet_3$, \ie the range of the Prandtl number $\rho$ and the Reynolds number $\nu$. We  consider {five} different ROM construction problems according to  the following setups: }

 \begin{itemize}
 \item[{\it i})] $D=30$ , $T=2$, $\zeta=0$, $\gamma=0$,   $\rho=0$ and $\nu=30$,%\remCH{Pourquoi mets-tu des accolades pour les 2 derniers parametres et pas aux autres?}
\item[{\it ii})]  identical to setup ({\it i}) but with the noise variance $\zeta^2$  set to induce a peak-to-signal-noise-ratio  around $26$,
\item[{\it iii})] identical to setup ({\it ii}) but  with an initial distribution whose  support  is a high-dimensional slice of thickness  $2\gamma=2\times10^{-3}$,
\item[{\it iv})] identical to setup ({\it iii}) but  with  longer trajectories ($T=5$)  and fewer observations ($D=10$),
\item[{\it v})] identical to setup ({\it iv}) but  with a  Prandtl number of $\rho=0.03$ and a Reynolds number in the interval $\nu\in  [30,300]$.
% \item setting {\it iv)},  same as  setting {\it iii)}, but  with  longer trajectories ($T=5$)  and fewer observations ($D=10$),
% \item setting {\it v)},  same as  setting {\it iv)}, but  with a dynamical coupling $\nu=\{0.03\}$, but a larger range for Prandtl number $\rho\in  [30,300]$.\\
 \end{itemize}
 The choice of $(T,D)=(2,30)$ and $T=(5,10)$ will be justified in  Section~\ref{sec:res}.

\subsection{ROMs and Sampling Algorithms} \label{sec:algos}
 We consider the two examples of reduced models exposed previously and their idealised\footnote{The term {\it idealised} refers to the fact that these ROMs commit an error only outside their approximation subspace. Of course they do not present any interest from a practical point of view since they require the computation of the high-dimensional states.} version noted with a star superscript, namely:
 
  \begin{itemize}
 
 \item \textit{ROM-1}, a POD-Galerkin approximation, presented  in Section~\ref{sec:POD};
 \item \textit{ROM-1$^\star$}, approximation  $\tilde x_t =\D\D^\intercal x_t$  for $t=1,...,T$  where  $\D$ is the parameter of \textit{ROM-1}, \ie the orthogonal projection on the approximation subspace of  \textit{ROM-1} defined by
\begin{align}\label{eq:S2} \underbrace{\textrm{im}(\D ) \times \cdots \times \textrm{im}(\D )}_{T\,\, \textrm{times}};\end{align} 
 %\footnote{
%The orthogonal projection   of $\x \in \Rr^{n \times T}$ on   subspace $\mathcal{S}_{2}$ minimises  the distance
%\begin{align*}
%  \textrm{dist}(\x, \mathcal{S}_2(\D)) &=\inf_{(x'_1,...,x'_T) \in  \mathcal{S}_2(\D)}\sum_{t=1}^T\|x_{t}-x_{t}'\|_2,\\
% &= \| \x-\D\D^\intercal \x  \|_F.
% \end{align*}
% }  
  
  %, \remCH{$\hat{\x}(\mathbf{u})= \mbox{projection de $x$ sur $\mathcal{S}(\mathbf{u})$}$? Ca permettrait de simplifier un peu la suite. Cela permettrait aussi de voir, pour chaque combinaison "sampling-methode de reduction", quelle est la part des degradations dues a une mauvaise identification d'un bon sous-espace d'aproximation $\cal{S}$ et celle due a des projections de type Galerkin/approximations lineaires. Attention, si tu decides de faire cela, il faut reintroduire ici la notation $\mathcal{S}(\mathbf{u})$ que j'ai eliminee du reste du manuscrit.}{   The mean distance between the $D$ trajectories which have generated the partial observations \remCH{de facon plus claire, celles generees a partir de $\px$, non?} and the approximation subspace $\mathcal{S}(\mathbf{u})$ (referred to as ``{lower bound}'' in the figures' legend), that is
% \begin{align}
%\mbox{mettre expression}\nonumber \\
%\nonumber
%\end{align}
%}\\
 \item \textit{ROM-2}, a low-rank linear approximation, presented in Section~\ref{sec:POP};
  \item \textit{ROM-2$^\star$},  approximation $\tilde x_{t}=\D x_{t-1}$ for $t=2,...,T$ and $\tilde x_1=x_1$  where  $\D$ is the parameter of \textit{ROM-2}, \ie  the orthogonal projection  on the approximation subspace  of \textit{ROM-2} defined by\footnote{
To see that the subspace defined in \eqref{eq:S1}, say $\mathcal{S}$, corresponds to the approximation subspace of {\it ROM-2}, we  remark  that the distance   of $\x \in \Rr^{n \times T}$ to this subspace is
\begin{align*}
\inf_{(x'_1,...,x'_T) \in  \mathcal{S}}\sum_{t=1}^T\|x_{t}-x_{t}'\|_2
 &=\inf_{(z_1,...,z_{T-1}) \in \Rr^{k \times (T-1)} }\sum_{t=2}^{T}\|x_{t}-\D \mathbf{w}_\D z_{t-1}\|_2,  \\
   &=\sum_{t=2}^{T}\|x_{t}-\D \mathbf{w}_\D \mathbf{w}_\D^\intercal \D^\dagger x_{t}\|_2,
 \end{align*}
and that this distance vanishes if  $x_{t}=\D \mathbf{w}_\D \mathbf{w}_\D^\intercal \D^\dagger x_{t}=\D x_{t-1}$ for $t=2,...,T$.
 } 
\begin{align}\label{eq:S1} \Rr^n\times \underbrace{\textrm{im}(\D \mathbf{w}_\D) \times \cdots \times \textrm{im}(\D \mathbf{w}_\D)}_{T-1\,\, \textrm{times}}. \end{align}
\end{itemize}
 
According to lower bounds in \eqref{eq:bornesError} and \eqref{eq:boundDMD},  {\it ROM-1}  and {\it ROM-2} will necessarily be less or as accurate as \textit{ROM-1$^\star$}  and \textit{ROM-2$^\star$}. The loss in accuracy between the ROMs and their idealised versions corresponds to the contribution of the error committed  by the ROMs inside their approximation subspaces.\\

%Eigenvectors of matrix $\mathbf{o}\mathbf{o}^\intercal\in \Rr^{n\times n}$ needed to build  \textit{ROM-1} are derived from the eigendecomposition of the smaller matrix $\mathbf{o}^\intercal\mathbf{o}\in \Rr^{TDN\times TDN}$, see \cite[Remark 1]{HeasHerzet16}.   \textit{ROM-2} is computed relying on SVD, see \cite[Algorithm 1]{HeasHerzet16}.\\
 
We assess  different sampling algorithms for building  {\it ROM-1},  {\it ROM-1$^\star$},  {\it ROM-2} and \textit{ROM-2$^\star$}. The parameter $\D$ of the ROMs are obtained using Proposition \ref{prop11} (resp. Proposition \ref{prop22}) for {\it ROM-1} and \textit{ROM-1$^\star$} (resp. for  {\it ROM-2} and \textit{ROM-2$^\star$})  with  a definition of matrices  $ \mathbf{a}$, $\mathbf{b}$ and $\mathbf{c}$  specific to the sampling algorithm. In the context of our SMC simulations, we observe that a number of particle  $N=40$ is reasonable. Indeed, increasing this number does not impact significantly the value of the inferred  ROM parameter. The sampling strategies are as follows.

%\remCH{reformuler cette partie pour mieux faire le lien avec le corps du texte}\remCH{NB: tu repetes a chaque item ci-dessous que le modele reduit est calcule a partir des Propositions 4.1 et 4.2. Ne serait-il pas plus judicieux de mentionner cela une fois pour toute dans la liste ci-dessus?}\\ 
 \begin{itemize}

 \item {\it Sampling  the target density  $p_\X$}.   ROMs are ideally built relying on samples drawn according to the (unknown) density $p_\X$.
 Matrices   $ \mathbf{a}$,~$\mathbf{b} ~\in~\Rr^{n \times (T-1)D}$ and $\mathbf{c}  \in \Rr^{n \times TD}$ are set in this case to
  \begin{align*}
 \mathbf{a} &= (  \x^{(1)}_{1:T-1}\cdots   \x^{(D)}_{1:T-1}),\\ 
 \mathbf{b}&= (  \x^{(1)}_{2:T}\cdots   \x^{(D)}_{2:T}) ,\\
   \mathbf{c}  &= ( \x^{(1)}_{1:T}\cdots  \x^{(D)}_{1:T}),
 \end{align*}  
 with $ \x^{(i)}_{t_1:t_2}=(x^{(i)}_{t_1} \cdots x^{(i)}_{t_2})$  where  $x^{(i)}_t$ is the hidden state which was used to generate observation $y^{(i)}_t$.
 % (referred to as ``{no uncertainty}'' in the figures' legend).\\

 \item {\it Sampling the proposed  data-enhanced surrogate density} $\ppx$. %\remCH{Je n'appelerais plus ca "posterior sampling" parce que finalement on echantillonne un version raffinee de $\tilde{p}(\x)$ definie comme $\hat{p}(\x)=\int_\mathbf{y} \hat{p}(\mathbf{x}\vert \mathbf{y}) p(\mathbf{y})$. Dire plutot: Sampling from the proposed data-enhanced surrogate density $\ppx$}.
   ROMs  are  built relying on a refined version of the surrogate $\spx$ defined in \eqref{eq:defppx}.  This density is approximated using MC and SMC techniques, as presented in  Sections~\ref{sec:mc} and~\ref{sec:smc}. More precisely, SMC samples are obtained by sequential importance sampling \cite{doucet2000sequential} with $\eta_1$ as the proposal distribution\footnote{In the case of a small noise variance $\zeta^2$, we slightly modify the density $\eta_1$ in \eqref{eq:recOrginalNon2} to avoid spreading  samples at initial time too far from observations. We  force  the initial condition samples to concentrate around the hyperplane $\mathbf{h}^\dagger y_1$ %, which corresponds to admissible points generating  observations $y_1^i$ in a noise-free context. To this aim, 
 by substituting sample $\tilde x_1$ by the orthogonal projection of 
$$ 
(\mathbf{i}_n-\mathbf{h}^\dagger\mathbf{h} )\tilde x_1 +\mathbf{h}^\dagger (y_1+\textrm{w}),
$$
on the subspace embedding $ \sparamSet_1$, where $\mathbf{h}^\dagger$ is the Moore-Penrose pseudo-inverse of $\mathbf{h}$.  Vector $\textrm{\w}\in \Rr^{n/2}$ is a realisation of a zero-mean and uncorrelated Gaussian  random variable whose component's variance is~$\zeta^2$.}.
Matrices $ \mathbf{a}$, $\mathbf{b}$ and $\mathbf{c}$ are  defined in Section~\ref{sec:POD} and Section~\ref{sec:POP}.

 \item {\it Sampling  the initial surrogate density} $\spx$. ROMs are built relying  on samples drawn according to $\spx$, \ie ignoring  observations. Matrices $ \mathbf{a}$, $\mathbf{b}  \in \Rr^{n \times (T-1)DN}$ and $\mathbf{c}  \in \Rr^{n \times TDN}$  are set in this case to
  \begin{align*}
 \mathbf{a} &= ( \XI^{(1,1)}_{1:T-1}\cdots  \XI^{(D,N)}_{1:T-1}),\\ 
 \mathbf{b}&= ( \XI^{(1,1)}_{2:T}\cdots  \XI^{(D,N)}_{2:T}) ,\\
  \mathbf{c}  &= (\XI^{(1,1)}_{1:T}\cdots \XI^{(D,N)}_{1:T}).
 \end{align*}  
 \item {\it Point estimates}.  ROMs are built relying on MMSE point estimates. Matrices $ \mathbf{a}$, $\mathbf{b}  \in \Rr^{n \times (T-1)D}$ and $\mathbf{c}  \in \Rr^{n \times TD}$ are set in this case to
 \begin{align*}
 \mathbf{a} &= ( \hat \XI^{(1)}_{1:T-1}\cdots \hat  \XI^{(D)}_{1:T-1}),\\ 
 \mathbf{b}&= ( \hat \XI^{(1)}_{2:T}\cdots  \hat \XI^{(D)}_{2:T}) ,\\
  \mathbf{c}  &= (\hat \XI^{(1)}_{1:T}\cdots \hat \XI^{(D)}_{1:T}),
 \end{align*}  
  with point estimates $ \hat \XI^{(i)}_{t_1:t_2}=(\hat \xi^{(i)}_{t_1} \cdots \hat \xi^{(i)}_{t_2})$   given for $i=1,...,D$  by  
 $$
\hat \xi^{(i)}_t = \sum_{j=1}^N w^{(i,j)} \xi^{(i,j)}_{t},
 $$
 for $t=1,...,T$.
  \end{itemize}

 %The performance of these algorithms for ROM construction  are given in terms of the $\ell_2$-norm  of the error between trajectories of the  original and reduced systems, \ie the average of the error norm $\parallel \x  -\tilde \x(\D)  \parallel_F$ over the $D$ trajectories which have generated the observations. %The performance of the different procedures are compared to the following benchmarks:\\
% \begin{itemize}
% \item  {\it No uncertainty}. Mean error  induced by a ROM built by sampling directly the prior $p_\X$, \ie knowing perfectly the $D$ trajectories which have generated the partial observations.
% \item  {\it Lower bound in \eqref{eq:boundErrorEstim}}. Mean distance   of the $D$ trajectories which have generated the partial observations to the approximation subspace. \\
 %\end{itemize}
 
 %\remCH{Je pense que l'on peu enlever l'itemize ci-dessus si on definit un modele reduit et un sampling supplementaires (cfr ci-dessus). Mais du coup, il faudrait sans rajouter les courbes correspondantes. }
 
% \remCH{il serait aussi interessant de montrer quelle erreur d'approximation on obtient avec les sous-espace calcule dans les ROMs lorsqu'on projette des realisations de $\spx$ ou $\ppx$. Cela nous donnerait une mesure de l' "epaisseur" de $\spx$ et $\ppx$ (parallele avec l'epaisseur de Kolmogorov)}
 
\subsection{Results and Discussion}\label{sec:res}

Figure~\ref{fig:4} and Figure~\ref{fig:4end} present the performance of the different sampling algorithms for building {\it  ROM-1},  \textit{ROM-1$^\star$},  {\it ROM-2} and   \textit{ROM-2$^\star$}. Figure~\ref{fig:4} and Figure~\ref{fig:4end} treat respectively  problem setups $i)$ to $iii)$ and  setups $iv)$ to $v)$.  The plots display the evolution of the average of the error norm $\parallel \x  -\tilde \x(\D)  \parallel_F$ over the $D$ trajectories which have generated the observations,  with respect to the ROM dimension $k$.    

We have set  $(T,D)=(2,30)$ in Figure~\ref{fig:4} and $(T,D)=(5,10)$ in Figure~\ref{fig:4end} to make the error norm comparable in  the two figures  for a dimension $k <50$.
 We  mention that  setting  $T=2$ (results displayed in Figure~\ref{fig:4}),  we obtain  that $\|x_2-\D x_1 \|_2=\|\x-\tilde \x (\D) \|_F$ for  {\it ROM-2}, implying that the distance to  the subspace  defined in \eqref{eq:S1} will necessarily be equal to the norm of the ROM error.  {\it ROM-2} will thus be in this case  equivalent to  \textit{ROM-2$^\star$} for any sampling algorithm. This setting simplifies the understanding and comparison of the different algorithms, as detailed below.   
Besides, for $T>2$ (Figure~\ref{fig:4end}) we can expect a difference of performance between {\it ROM-2} and  \textit{ROM-2$^\star$} , and in particular, the more non-linear the $f_t$'s in the dynamical model \eqref{eq:model_init}, the more the difference of performance.
  Nevertheless, in our experiments, we will observe that this difference remains reasonable,  showing that inequalities \eqref{eq:bornesError} and \eqref{eq:boundDMD} are almost equalities ($c\simeq1$), and in consequence this will provide an experimental justification to  the strategy of bounding (rather than evaluating precisely) the objective function.
 For legibility purposes, we will display the performance of  \textit{ROM-1$^\star$} and  \textit{ROM-2$^\star$} only for the algorithm   sampling the target density. \\
  
%\remCH{Tu discutes les resultats sans jamais avoir fait une seule fois reference aux figures. Il faut d'abord presenter les figures et dire a quoi elles correspondent sinon le reste de la discussion devient peu comprehensible.}  
We observe  in our experiments that, for any problem setup and any ROM, sampling the {\it initial surrogate}  $\spx$  leads to the poorest peformance. The ROMs built from {\it point estimates} yields to a slight enhancement. Moreover, except in the case of setup~$v)$ and its strong non-linearities, the {\it proposed data-enhanced surrogate}  leads to the best approximation accuracy. In what follows, we discuss in details this general analysis.  \\
 
Let us begin by some comments on the behaviour of the  algorithm   sampling the {\it target density} $p_\X$ for construction of {\it  ROM-1}, {\it  ROM-2} and their  idealised versions.     
As expected  for  setups $i)$ and $ii)$ where $T=2$, the error of  {\it ROM-2}  %\remCH{Les appellations ROM-2 et ROM-1 n'apparaissent pas sur les figures (ou tu utilises "Low rank approx" et "Galerkin projection"). Sont elles necessaires, cela brouille plus le message qu'autre chose je trouve.} 
  %, \ie of a low-rank DMD approximation, 
vanishes\footnote{The computation of an eigendecomposition or the use of SVD induces a machine precision around $1e-5$ for trajectories computed with {\it ROM-1} or {\it ROM-2}.}  for a subspace of dimensionality  above the initial condition dimension, \ie  for $k\ge 10$.  {\it ROM-1} is  less accurate because  the error  vanishes in the best case  for $k\ge 20$  in the case of a projection of the trajectory on  the subspace defined in \eqref{eq:S2} (while it can vanish for $k\ge 10$   for a projection on $\mathcal{S}_1$).  For setup $iii)$, the initial conditions are not embedded anymore in a $10$-dimensional subspace, but in a slice of thickness $2\gamma$ and of dimensionality $10+40=50$ around this subspace. As expected, we verify that the  error is equal to zero for $k \ge \min(50,D)=30$ (resp. $k \ge \min(100,2D)=60)$  in the case of  {\it ROM-2}  (resp. {\it ROM-1}). In practice, we note that the error vanishes even around  $k \ge 25 $ (resp. $k \ge 50)$.  The last two settings, \ie setups $iv)$ and $v)$, imply longer sequences $T=5$. In this situation, the trajectories could be approximated in a worst-case scenario  (scenario of linear independence of states at different times of the trajectory) using a subspace of dimension  $(T-1)D=40$  for {\it ROM-2}  (resp. of dimension  $TD=50$ for {\it ROM-1}).  We observe however that this pessimistic scenario does not occur in practice, especially if non-linearities are moderated (setup $iv) )$. Indeed, between $30$ and $40$ components (resp. between $35$ and $40$ components) are sufficient to obtain a zero approximation error for {\it ROM-2}  (resp. {\it ROM-1}). \\

We now compare the performance of algorithms which do not rely on the knowledge of $p_\X$, namely the algorithms sampling  the {\it initial surrogate}, the {\it proposed data-enhanced surrogate} and the algorithm based on {\it point estimates}.  

First, for  setups $i)$ and $ii)$,  since the {\it initial surrogate} and  the {\it proposed data-enhanced surrogate} are defined for an  initial condition of dimensionality twice bigger, we verify that  {\it ROM-2} (resp.   {\it ROM-1}) built by sampling the   {\it initial surrogate} or the {\it proposed data-enhanced surrogate} cancel the error  for $k\ge 20$ (resp. $k\ge 40$). We can observe that a ROM built with {\it point estimates} is  slightly less accurate independently of the dimension $k$, and in particular for $k\ge 20$ by a factor $5$. This error saturation effect is likely to be  the consequence of a large variance of $\spx$ in certain directions of the kernel of  $\mathbf{h}$. In other words, this non-reducible error is possibly due to the fact the method ignores that a single point estimates is insufficient to represent variable $\X$, along some of the non-observed directions. 
 % \remCH{Mentionnes tu quelque part le nombre $N$ de particules que tu consideres? Quel est l'impact du nombre de particules sur les performances finales?}
 
Second, we observe a moderate loss of accuracy  for setup $ii)$, which attests that the data-based algorithms seem to be robust to moderate observation noise. %\remCH{le bruit est ridiculement petit non ($\zeta^2=10^{-10}$)? Est-ce suffisant pour clamer la resistance au bruit?}  
    
Third,  the target distribution defined in setup $iii)$ (taking the form of a high-dimensional slice) turns out to slightly increase  the error above $k\ge 20$ by a factor $5$ (resp. $2.5$) when  sampling  the {\it initial surrogate} or  the {\it proposed data-enhanced surrogate}  (resp. for {\it point estimates}).  %Although the present model reduction minimises an average square error rather than adopting a worst-case perspective, 
This result can be interpreted as the fact that the algorithms are robust to the reduction of trajectories  which do not necessarily belong to a subspace, but which are moderately distant  from it. 
% in the light of the theoretical analysis led in  \cite{HerzetHeasDremeau16}. \remCH{Je ne sais pas trop ce que tu veux dire mais si tu gardes la phrases %precedente, alors il faut un peu preciser la nature de l'analyse que l'on peut faire a partir des resultats de \cite{HerzetHeasDremeau16}. }
  
  Fourth, we observe in setup $iv)$, that sampling the {\it proposed data-enhanced surrogate}   or using {\it point estimates} induces only a slight deterioration of  performances  when  compared to the algorithm sampling the {\it target density}.  On the contrary,  while being reasonable  for {\it ROM-1}, sampling the {\it initial surrogate}  yields dramatic results for {\it ROM-2}. This effect can be easily understood: in the case of a low-rank linear approximation, increasing the dimension $k$  is not sufficient to obtain a gain in  performance; in particular,  over-estimating the eigenvalues of matrix $\D^\star$ induces by construction of {\it ROM-2}  an exponential increase of the approximation  error. %\remCH{Je ne suis pas sur de bien comprendre ton analyse de ce qui se passe}
 
   Finally,  for  setup~$v)$,  this unstable behaviour affects   {\it ROM-2} for any of the algorithms.  On the contrary, {\it ROM-1}, \ie POD-Galerkin approximation, seems to be nearly insensitive to the presence of strong non-linearities in \eqref{eq:model_init}.
\begin{figure}[t!]
\centering
\begin{tabular}{c|c}
\includegraphics[width=0.5\columnwidth]{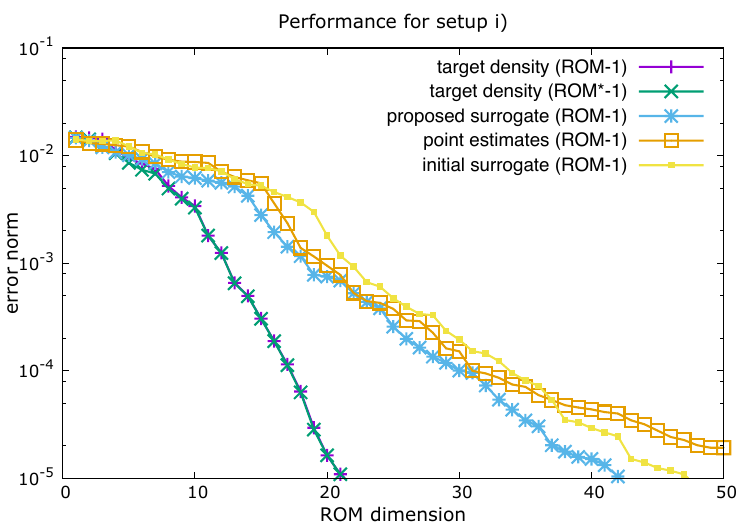}&\includegraphics[width=0.5\columnwidth]{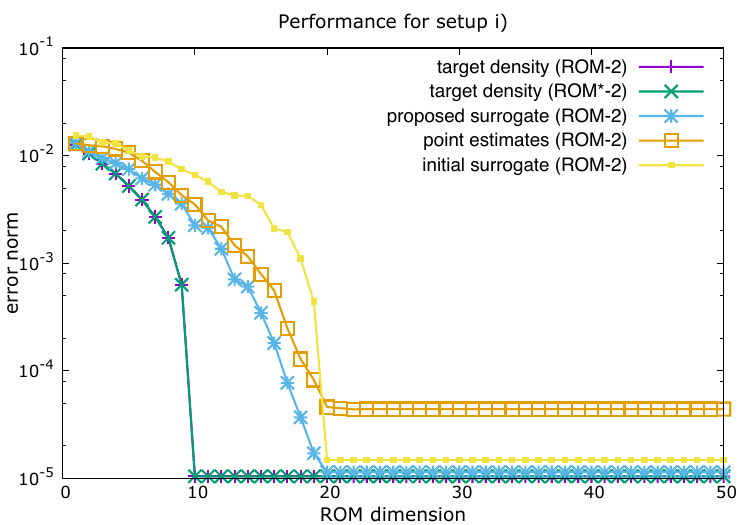}\\
\includegraphics[width=0.5\columnwidth]{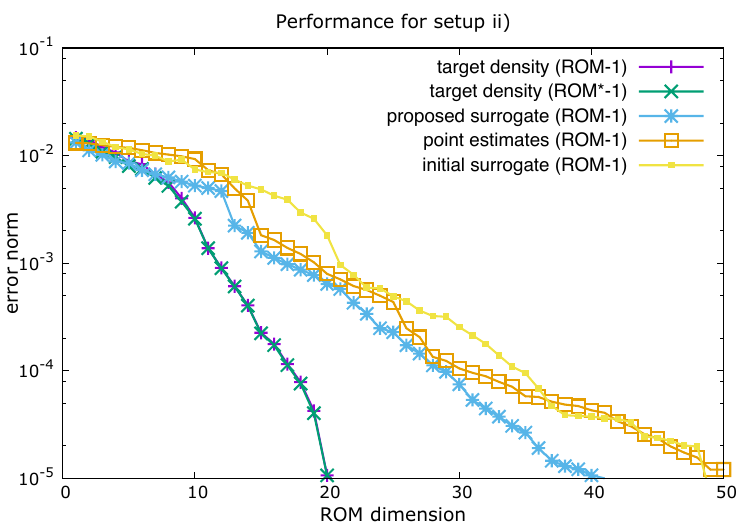}&\includegraphics[width=0.5\columnwidth]{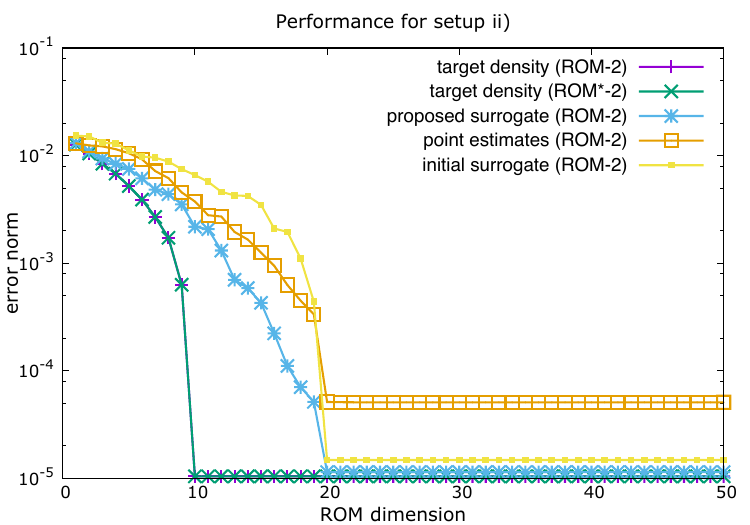}\\
\includegraphics[width=0.5\columnwidth]{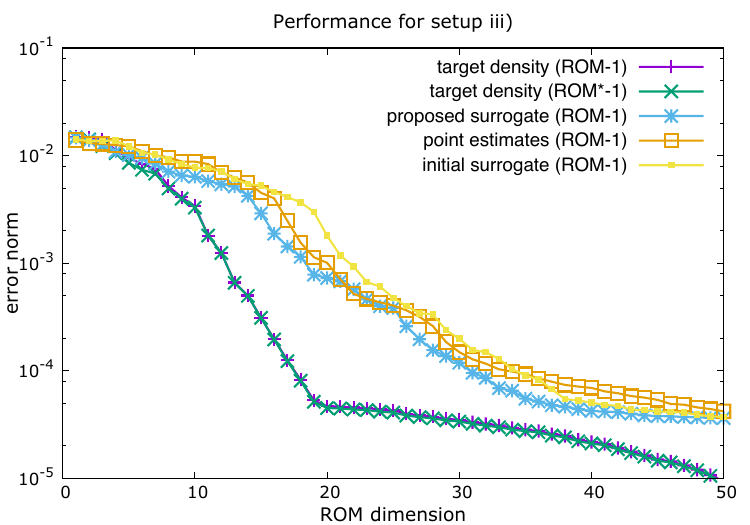}&\includegraphics[width=0.5\columnwidth]{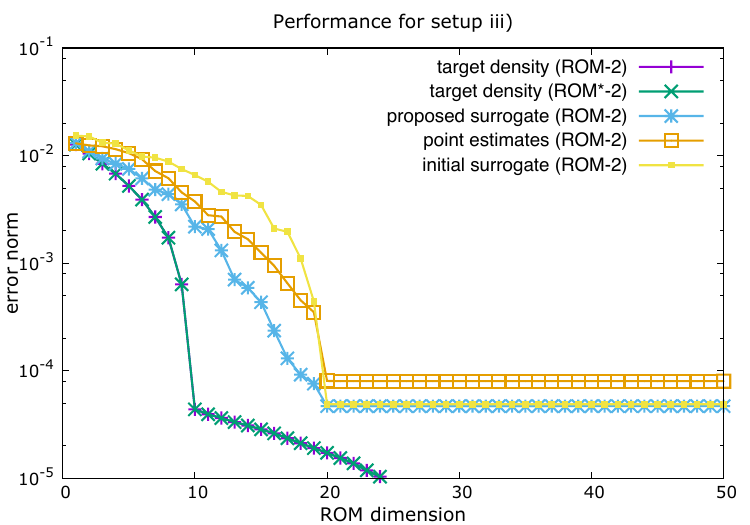}\\
\end{tabular}
\caption{\small 
{
%\remCH{Je ne comprends pas le comportement de la courbe jaune entre les figures de gauche et de droite: en principe, je m'attendrais a ce que la POD atteignent une meilleure erreur d'approximation etant donne que le sous-espace d'approximation est optimise selon ce critere, non?} 
%\remCH{Il faut mettre des marqueurs aux courbes, sinon elles deviennent illisibles lorsqu'on imprime le papier.}
Algorithms performances for construction of   {\it ROM-1} and  \textit{ROM-1$^\star$} (on the left) and {\it ROM-2} and  \textit{ROM-2$^\star$}  (on the right)  for setup $i)$ (above), $ii)$ (middle) and $iii)$ (below).  See details in Sections~\ref{sec:data} and~\ref{sec:algos} }
\vspace{-0.3cm}\label{fig:4}}
\end{figure}

\begin{figure}[t!]
\centering
\begin{tabular}{c|c}
\includegraphics[width=0.5\columnwidth]{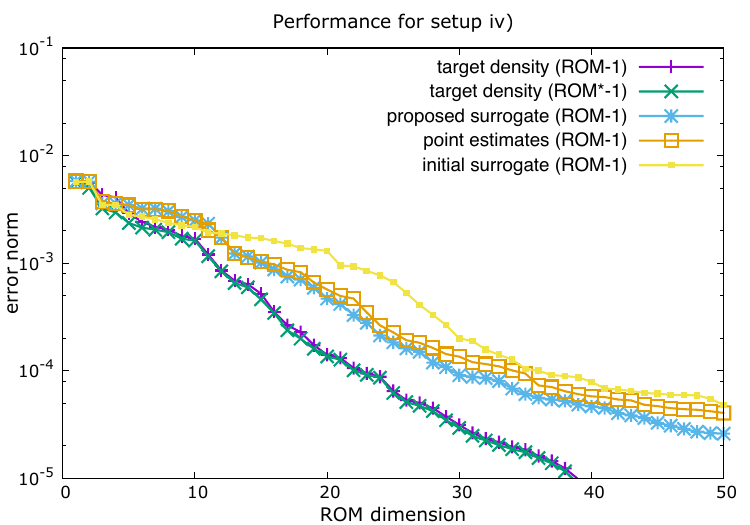}&\includegraphics[width=0.5\columnwidth]{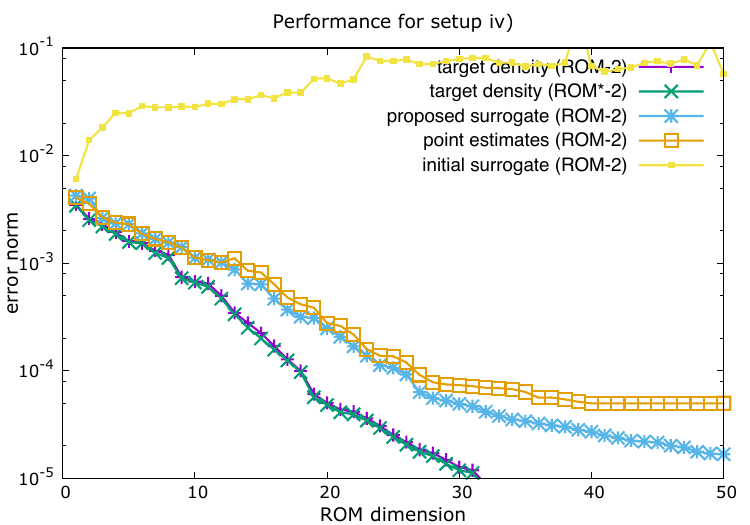}\\
\includegraphics[width=0.5\columnwidth]{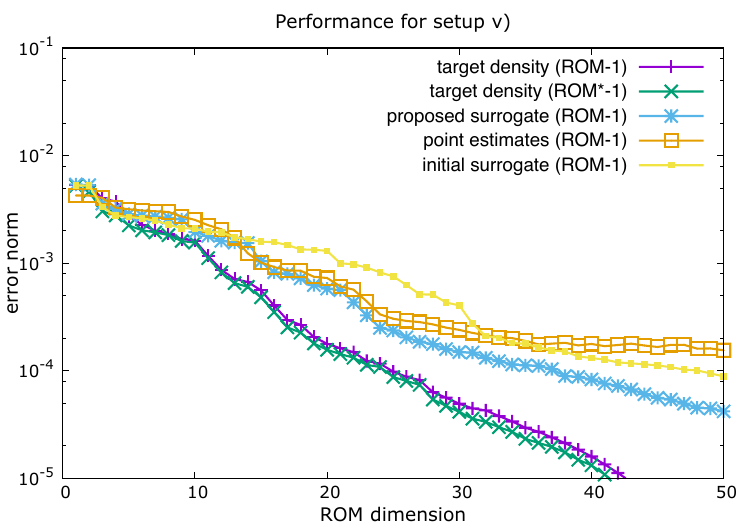}&\includegraphics[width=0.5\columnwidth]{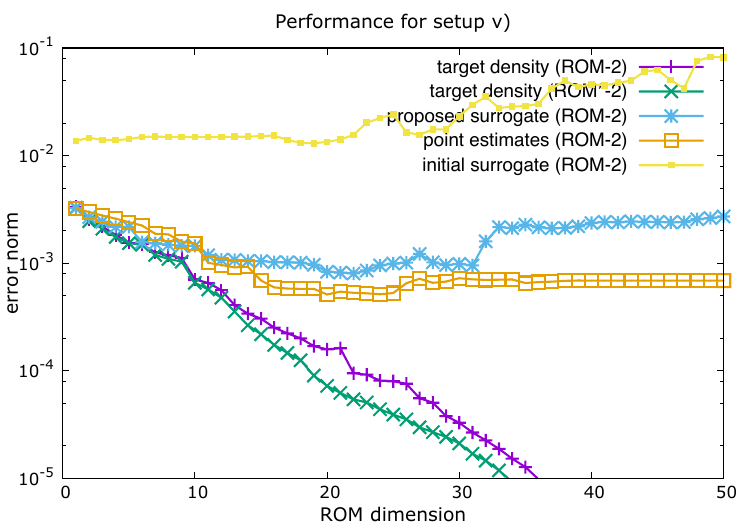}
\end{tabular}
\caption{\small {
Algorithms performances for construction of   {\it ROM-1} and  \textit{ROM-1$^\star$} (on the left) and {\it ROM-2} and  \textit{ROM-2$^\star$}  (on the right)  for setup $iv)$ (above) and $v)$ (below).  See details in Sections~\ref{sec:data} and~\ref{sec:algos}}
\vspace{-0.3cm}\label{fig:4end}}
\end{figure}

\section{Conclusions}
%\remCH{Reformuler l'intro pour prendre en compte la nouvelle version du papier?}
We have proposed a  general framework for the construction of ROMs  when the distribution of the trajectories of a dynamical system  is imperfectly known. This work assumes that we have  the following two sources of information at our disposal: 1) an initial surrogate density characterising  the trajectories of the system of interest; 2) a set of incomplete observations on the target trajectories obeying a {known} conditional model.

The ROM construction consists in  the minimisation of the  expectation of  the norm of the  error  between the true and reduced trajectories. The expectation relies on a data-enhanced surrogate density obtained in a Bayesian setting combining the initial  surrogate density and the conditional observation model. 
We show, that under mild conditions,  the proposed data-enhanced surrogate is a better approximation than  the initial surrogate in term of  the Kullback-Leibler distance to the target density.

We stress the need of approximations to efficiently solve this problem and propose in this context tractable solvers.  In particular, we use  MC and SMC techniques to characterise our data-enhanced surrogate and propose  implementations based on the minimisation of a bound on the objective function.   We illustrate how the proposed methodology  particularises to two different families of ROMs. We  show that these particularisations can be seen as generalisations, in a probabilistic framework, of well-known ROM techniques, namely POD and low-rank DMD. We also  show that they can be seen as generalisations of standard ROM constructions based on point estimates.

 A numerical evaluation, led in the context of the reduction of a geophysical model, reveals that  the proposed methodology may enhance state of the art.

\appendix

\section{Low-Rank Linear Approximation}%\remCH{je ne comprends pas certains passages de cette preuve: a rediscuter ensemble avant la soumission finale.}\\

\subsection{Error Bounds}\label{app:2} 
We hereafter show  that the error norm for a  low-rank linear approximation  of \eqref{eq:model_init} can be bounded  as presented in Section~\ref{sec:POP}.  
On the one hand, to obtain the lower bound in \eqref{eq:boundDMD}, we notice that according to \eqref{eq:S1}, we have $\tilde x_t \in \textrm{im}(\D \mathbf{w}_\D)$  for $t\ge2$, so that
each term contributing to the  error norm  $ \parallel \x  -\tilde \x(\D)  \parallel^2_F=\sum_{t=2}^{T}\| x_t-\tilde x_t(\D)\|_2^2 $ can be decomposed into two orthogonal components  $$  x_t-\tilde x_t(\D) = \underbrace{\D \mathbf{w}_\D \mathbf{w}_\D^\intercal\D^\dagger x_t -\tilde x_t(\D)}_{\in \, \textrm{im}(\D \mathbf{w}_\D)}+\underbrace{x_t-\D \mathbf{w}_\D \mathbf{w}_\D^\intercal\D^\dagger x_t}_{\in \,  \textrm{im}(\D \mathbf{w}_\D)^\perp}. $$ 
This implies that $$\sum_{t=2}^T\|x_t-\D \mathbf{w}_\D \mathbf{w}_\D^\intercal\D^\dagger x_t \|^2_2=\sum_{t=2}^{T}\|x_{t}-\D x_{t-1}\|^2_2 \le  \parallel \x  -\tilde \x(\D)  \parallel^2_F. $$

On the other hand, since elements of the set \eqref{eq:adminstePOP} are such that $\|\D\|_{2,2} \le \lambda<\infty$, the following result shows  the upper bound in \eqref{eq:boundDMD}. % can be obtained  under mild assumptions. 
\begin{lemma}
%Assume there exists $\lambda$ such that  $\|\D\|_{2,2}< \lambda. $ 
$ %\begin{align*}
% \mathbf{e}_{pod}(\x ,\tilde \x(\D))
  \parallel \x  -\tilde \x(\D)  \parallel^2_F \le   c \, \sum_{t=2}^{T}\|x_{t}-\D x_{t-1}\|^2_2 , $
  %\end{align*}
with $c=\max_{t\in\{2,...,T\}}\{\sum_{k=t}^T (1+4(k-1)) \|\D\|_{2,2}^{2(k-t)}\} $. 
 \end{lemma}
\proof{
Since $\tilde x_t= \D \tilde x_{t-1},$ we have  
 \begin{align*}
 \parallel \x  -\tilde \x(\D)  \parallel^2_F&= \sum_{t=2}^{T}\| x_{t}-\D^{t-1}x_{1}\|_2^2.
 \end{align*}
Using the triangular inequality and the definition of the induced $\ell_2$-norm, each term contribution in this sum can be bounded as follows
\begin{align*}
\| x_{t}-\D^{t-1}x_{1}\|_2& \le \sum_{\ell=2}^{t} \|\D^{\ell-1}x_{t-\ell+1}-\D^{\ell-2}x_{t-\ell+2} \|_2, \\
%&\|\D^{t-1}x_{1}-\D^{t-2}x_{2} \|_2+\|\D^{t-2}x_{2}-\D^{t-3}x_{3} \|_2  \\
%&+ \cdots +\|\D^{1}x_{t-1}-x_{t} \|_2,\\
&\le  \sum_{\ell=2}^{t} \|\D^{\ell-2}\|_{2,2} \|\D x_{t-\ell+1}-x_{t-\ell+2} \|_2, \\
%&\le \|\D^{t-2}\|_{2,2}\|\D x_{1}-x_{2}\|_2 +  \|\D^{t-3}\|_{2,2}\|\D x_{2}-x_{3}\|_2 \\
%&+ \cdots + \|\D x_{t-1}-x_{t}\|_2,\\
&= \sum_{k=2}^t  \|\D^{t-k}\|_{2,2} \|\D x_{k-1}-x_{k}\|_2.
\end{align*}
Therefore, expanding the square of this sum, we obtain 
\begin{align*}
\| x_{t}-\D^{t-1}x_{1}\|^2_2&\le  \sum_{k=2}^t  \|\D^{t-k}\|^2_{2,2} \|\D x_{k-1}-x_{k}\|^2_2\\
&+2 \sum_{i,j =2| i \neq j}^t \|\D^{t-i}\|_{2,2} \|\D x_{i-1}-x_{i}\|_2 \|\D^{t-j}\|_{2,2} \|\D x_{j-1}-x_{j}\|_2,\\
&\le  \sum_{k=2}^t  \|\D^{t-k}\|^2_{2,2} \|\D x_{k-1}-x_{k}\|^2_2\\
&+2 \sum_{i,j =2| i \neq j}^t \max\{\|\D^{t-i}\|^2_{2,2} \|\D x_{i-1}-x_{i}\|_2^2, \|\D^{t-j}\|^2_{2,2} \|\D x_{j-1}-x_{j}\|_2^2\},\\
&\le  \sum_{k=2}^t  \|\D^{t-k}\|^2_{2,2} \|\D x_{k-1}-x_{k}\|^2_2\\
&+2 \sum_{i,j =2| i \neq j}^t  \|\D^{t-i}\|^2_{2,2} \|\D x_{i-1}-x_{i}\|_2^2 + \|\D^{t-j}\|^2_{2,2} \|\D x_{j-1}-x_{j}\|_2^2,\\
&\le  \sum_{k=2}^t (1+4(t-1)) \|\D^{t-k}\|^2_{2,2} \|\D x_{k-1}-x_{k}\|^2_2.\\
\end{align*}
In consequence, we conclude remarking that 
\begin{align*}
 \parallel \x  -\tilde \x(\D)  \parallel^2_F&\le \sum_{t=2}^{T}\sum_{k=2}^t (1+4(t-1)) \|\D^{t-k}\|^2_{2,2} \|\D x_{k-1}-x_{k}\|^2_2,\\
&= \sum_{t=2}^T  \sum_{k=t}^{T} (1+4(k-1)) \|\D^{k-t}\|^2_{2,2} \|\D x_{t-1}-x_{t}\|^2_2,\\
&\le c\,  \parallel \x_{2:T} -\D \x_{1:T-1} \parallel^2_F,
\end{align*}
with $c=\max_{t\in\{2,...,T\}}\{\sum_{k=t}^T (1+4(k-1)) \|\D^{k-t}\|^2_{2,2}\} $ and where the equality has been obtained by inverting the two sums. $\square$
}

\subsection{Finite Norm of Minimisers}\label{app:3}
{
Let $J_1(\D)=\sum_{t=2}^{T}\|x_{t}-\D x_{t-1}\|^2_2$,  $J_2(\D)= \| \x-\tilde \x(\D)  \|^2_F$ and  $\mathcal{U}_k=\{  \D \in \Rr^{n\times n}| \textrm{rank}(\D) \le k \}$.
 \begin{lemma}Minimisers of $J_1$ and $J_2$ over the domain  $\mathcal{U}_k$ have a finite  norm. \end{lemma}
 
%  We have $$ \argmin_{\D \in \mathcal{U}_k } J(\D) = \argmin_{\D \in \mathcal{U}} J(\D),  $$ with   $\mathcal{U}_k=\{  \D \in \Rr^{n\times n}| \textrm{rank}(\D) \le k \}$ and where  $J(\D)$ is defined in \eqref{eq:upperBoundPOP} and $\mathcal{U}$  in \eqref{eq:adminstePOP} with $\lambda < \infty$. 
\proof{
First notice that the objective functions are not infinite on all the optimisation domain (\eg $J_i(0) <\infty$ for $i=1,2$).  Next, let    $\mathcal{U}$ be defined  in \eqref{eq:adminstePOP} with $\lambda < \infty$ and  $$\mathcal{U}^\infty_k=\{  \D \in \Rr^{n\times n}| \textrm{rank}(\D) \le k,  \|\D\|_{2,2} =\infty \}.$$  We have   
 \begin{align*}
 \argmin_{\D \in \mathcal{U}_k } J_i(\D) \stackrel{(a)}{=} \argmin_{\D \in \mathcal{U} \cup \mathcal{U}^\infty_k } J_i(\D)\stackrel{(b)}{=} \argmin_{\D \in \mathcal{U} \cup (\mathcal{U}_{k'}\setminus \mathcal{U}^\infty_{k'}) } J_i(\D)\stackrel{(c)}{=} \argmin_{\D \in \mathcal{U}} J_i(\D) ,\quad \textrm{for }\quad i=1,2,
 \end{align*}
 with $k' <k$. Equality {\it (a)} follows from the decomposition $\mathcal{U}_k = \mathcal{U} \cup \mathcal{U}^\infty_k $. Equality {\it (b)}  is deduced from  the two  following facts.  For $\D \in \mathcal{U}^\infty_k$, let $\mathbf{v}^\infty_\D$ denote  the matrix whose columns are the right singular vectors of $\D$ associated to infinite singular values.   If $\D$   is such that ${x}_{t-1} \in \textrm{im} (\mathbf{v}^\infty_\D) $  for at least one of the indices $t=2,\cdots, T$ (resp. ${x}_{1} \in \textrm{im} (\mathbf{v}^\infty_\D) $),  then $J_1(\D)=\infty$ (resp. $J_2(\D)=\infty$) implying that $\D$ is not a minimiser. On the other hand, if  $\D$   is such that ${x}_{t-1} \notin \textrm{im} (\mathbf{v}^\infty_\D)$ for any of the indices  $t=2,\cdots, T$ (resp. ${x}_{1} \notin \textrm{im} (\mathbf{v}^\infty_\D)$),  then there exists  $\D' \in  (\mathcal{U}_{k'}\setminus \mathcal{U}^\infty_{k'}) $ such that $J_1(\D)=J_1(\D')$ (resp. $J_2(\D)=J_2(\D')$). Equality {\it (c)}  follows from the inclusion $(\mathcal{U}_{k'}\setminus \mathcal{U}^\infty_{k'}) \subset \mathcal{U}$.  $\square$
 }}
\section*{Compliance with Ethical Standards}{
%This work was
%supported by the ``Agence Nationale de la Recherche" (ANR). % through the GERONIMO project.
The  authors state that there is no conflict of interest. }

% BibTeX users please use one of
%\bibliographystyle{spbasic}      % basic style, author-year citations
\bibliographystyle{spmpsci}      % mathematics and physical sciences
\bibliography{./bibtex,bibtexDMD,cherzet,group-15332,Her_biblio}   % name your BibTeX data base

\begin{thebibliography}{10}
\providecommand{\url}[1]{{#1}}
\providecommand{\urlprefix}{URL }
\expandafter\ifx\csname urlstyle\endcsname\relax
  \providecommand{\doi}[1]{DOI~\discretionary{}{}{}#1}\else
  \providecommand{\doi}{DOI~\discretionary{}{}{}\begingroup
  \urlstyle{rm}\Url}\fi

\bibitem{2015arXiv151106196A}
{Agapiou}, S., {Papaspiliopoulos}, O., {Sanz-Alonso}, D., {Stuart}, A.M.:
  {Importance Sampling: Computational Complexity and Intrinsic Dimension}.
\newblock ArXiv e-print 1511.06196  (2015)

\bibitem{Antoulas2005Overview}
Antoulas, A.C.: An overview of approximation methods for large-scale dynamical
  systems.
\newblock Annual Reviews in Control \textbf{29}(2), 181--190 (2005)

\bibitem{Barrault2004667}
Barrault, M., Maday, Y., Nguyen, N.C., Patera, A.T.: An empirical interpolation
  method: application to efficient reduced-basis discretization of partial
  differential equations.
\newblock Comptes Rendus Mathematique \textbf{339}(9), 667 -- 672 (2004)

\bibitem{chandrasekhar2013hydrodynamic}
Chandrasekhar, S.: Hydrodynamic and hydromagnetic stability.
\newblock Courier Corporation (2013)

\bibitem{chaturantabut2010nonlinear}
Chaturantabut, S., Sorensen, D.C.: Nonlinear model reduction via discrete
  empirical interpolation.
\newblock SIAM Journal on Scientific Computing \textbf{32}(5), 2737--2764
  (2010)

\bibitem{Chen12}
Chen, K.K., Tu, J.H., Rowley, C.W.: Variants of dynamic mode decomposition:
  boundary condition, koopman, and fourier analyses.
\newblock Journal of nonlinear science \textbf{22}(6), 887--915 (2012)

\bibitem{Ciarlet02}
Ciarlet, P.: The Finite Element Method for Elliptic Problems.
\newblock Society for Industrial and Applied Mathematics (2002)

\bibitem{2015arXiv150206797C}
{Cohen}, A., {Devore}, R.: {Approximation of high-dimensional parametric
  {PDE}s}.
\newblock ArXiv e-print 1502.06797  (2015)

\bibitem{crisan2002survey}
Crisan, D., Doucet, A.: A survey of convergence results on particle filtering
  methods for practitioners.
\newblock IEEE Transactions on signal processing \textbf{50}(3), 736--746
  (2002)

\bibitem{CuiEtAl2014}
{Cui}, T., {Martin}, J., {Marzouk}, Y.M., {Solonen}, A., {Spantini}, A.:
  {Likelihood-informed dimension reduction for nonlinear inverse problems}.
\newblock Inverse Problems \textbf{30}(11), 114015 (2014)

\bibitem{CuiMarzoukWillcox2014}
{Cui}, T., {Marzouk}, Y.M., {Willcox}, K.E.: {Data-driven model reduction for
  the Bayesian solution of inverse problems}.
\newblock International Journal for Numerical Methods in Engineering
  \textbf{102}, 966--990 (2015)

\bibitem{doucet2000sequential}
Doucet, A., Godsill, S., Andrieu, C.: On sequential monte carlo sampling
  methods for bayesian filtering.
\newblock Statistics and computing \textbf{10}(3), 197--208 (2000)

\bibitem{Everson1995KarhunenLoeve}
Everson, R., Sirovich, L.: Karhunen-lo\`{e}ve procedure for gappy data.
\newblock J. Opt. Soc. Am. A \textbf{12}(8), 1657--1664 (1995)

\bibitem{ZAMM:ZAMM19830630105}
Fink, J.P., Rheinboldt, W.C.: On the error behavior of the reduced basis
  technique for nonlinear finite element approximations.
\newblock ZAMM - Journal of Applied Mathematics and Mechanics \textbf{63}(1),
  21--28 (1983)

\bibitem{Gunes2006Gappy}
Gunes, H., Sirisup, S., Karniadakis, G.E.: Gappy data: To krig or not to krig?
\newblock Journal of Computational Physics \textbf{212}(1), 358--382 (2006)

\bibitem{Hasselmann88}
Hasselmann, K.: Pips and pops: The reduction of complex dynamical systems using
  principal interaction and oscillation patterns.
\newblock Journal of Geophysical Research: Atmospheres \textbf{93}(D9),
  11,015--11,021 (1988)

\bibitem{HeasHerzet16}
{H{\'e}as}, P., {Herzet}, C.: {Low-rank Approximation and Dynamic Mode
  Decomposition}.
\newblock ArXiv e-print 1610.02962  (2016)

\bibitem{HerzetHeasDremeau16}
{Herzet}, C., {Dr{\'e}meau}, A., {H{\'e}as}, P.: {Model Reduction from Partial
  Observations}.
\newblock ArXiv e-print 1609.08821  (2016)

\bibitem{9780511622700}
Holmes, P., Lumley, J.L., Berkooz, G.: Turbulence, Coherent Structures,
  Dynamical Systems and Symmetry.
\newblock Cambridge University Press (1996).
\newblock Cambridge Books Online

\bibitem{jolliffe2002principal}
Jolliffe, I.: Principal Component Analysis.
\newblock Springer Series in Statistics. Springer (2002)

\bibitem{Jovanovic12}
Jovanovic, M., Schmid, P., Nichols, J.: Low-rank and sparse dynamic mode
  decomposition.
\newblock Center for Turbulence Research Annual Research Briefs pp. 139--152
  (2012)

\bibitem{Kwasniok96}
Kwasniok, F.: The reduction of complex dynamical systems using principal
  interaction patterns.
\newblock Phys. D \textbf{92}(1-2), 28--60 (1996)

\bibitem{Lorenz63}
{Lorenz}, E.N.: {Deterministic Nonperiodic Flow.}
\newblock Journal of Atmospheric Sciences \textbf{20}, 130--148 (1963)

\bibitem{NME:NME4747}
Maday, Y., Patera, A.T., Penn, J.D., Yano, M.: A parameterized-background
  data-weak approach to variational data assimilation: formulation, analysis,
  and application to acoustics.
\newblock International Journal for Numerical Methods in Engineering
  \textbf{102}(5), 933--965 (2015)

\bibitem{peherstorfer2016data}
Peherstorfer, B., Willcox, K.: Data-driven operator inference for nonintrusive
  projection-based model reduction.
\newblock Computer Methods in Applied Mechanics and Engineering \textbf{306},
  196--215 (2016)

\bibitem{quarteroni2015reduced}
Quarteroni, A., Manzoni, A., Negri, F.: Reduced basis methods for partial
  differential equations: an introduction, vol.~92.
\newblock Springer (2015)

\bibitem{Quarteroni2011Certified}
Quarteroni, A., Rozza, G., Manzoni, A.: Certified reduced basis approximation
  for parametrized partial differential equations and applications.
\newblock Journal of Mathematics in Industry \textbf{1}(1), 1--49 (2011)

\bibitem{Sirovich87}
{Sirovich}, L.: {Turbulence and the dynamics of coherent structures.}
\newblock Quarterly of Applied Mathematics \textbf{45}, 561--571 (1987)

\bibitem{SpantiniEtAl2015}
{Spantini}, A., {Solonen}, A., {Cui}, T., {Martin}, J., {Tenorio}, L.,
  {Marzouk}, Y.: {Optimal low-rank approximations of Bayesian linear inverse
  problems}.
\newblock ArXiv e-prints  (2014)

\end{thebibliography}

%
%% Non-BibTeX users please use
%\begin{thebibliography}{}
%%
%% and use \bibitem to create references. Consult the Instructions
%% for authors for reference list style.
%%
%\bibitem{RefJ}
%% Format for Journal Reference
%Author, Article title, Journal, Volume, page numbers (year)
%% Format for books
%\bibitem{RefB}
%Author, Book title, page numbers. Publisher, place (year)
%% etc
%\end{thebibliography}

\end{document}